\documentclass[nofootinbib,showpacs,preprint,preprintnumbers,amsmath,amssymb]{revtex4-1}
\pdfoutput=1
\usepackage{bm}
\usepackage{amsmath}
\usepackage{amsfonts}
\usepackage{amssymb}
\usepackage{float}
\usepackage{color}
\usepackage{graphicx}
\usepackage{graphics}
\usepackage{hyperref}
\usepackage{adjustbox,array}
\usepackage{enumitem} 
\hypersetup{}
\usepackage[utf8]{inputenc} 
\usepackage{epstopdf}
\usepackage{multirow}
\usepackage{slashed}
\usepackage{booktabs}
\usepackage{cancel}
\usepackage{mathrsfs}
\usepackage{csquotes}
\usepackage{mathtools}
\usepackage{rotating}
\usepackage{afterpage}

\DeclareUnicodeCharacter{2217}{}
\DeclareUnicodeCharacter{2212}{}

\definecolor{rosso}{cmyk}{0,1,1,0.4}
\definecolor{rossos}{cmyk}{0,1,1,0.55}
\definecolor{rossoc}{cmyk}{0,1,1,0.2}
\definecolor{blu}{cmyk}{1,1,0,0.3}
\definecolor{blus}{cmyk}{1,1,0,0.6}
\definecolor{bluc}{cmyk}{1,1,0,0.1}
\definecolor{verde}{cmyk}{0.92,0,0.59,0.25}
\definecolor{verdec}{cmyk}{0.92,0,0.59,0.15}
\definecolor{verdes}{cmyk}{0.92,0,0.59,0.4}

\AtBeginDocument{
\heavyrulewidth=.08em
\lightrulewidth=.05em
\cmidrulewidth=.03em
\belowrulesep=.65ex
\belowbottomsep=0pt
\aboverulesep=.4ex
\abovetopsep=0pt
\cmidrulesep=\doublerulesep
\cmidrulekern=.5em
\defaultaddspace=.5em
}

\newcommand{\orcid}[1]{\hspace{1mm}\href{https://orcid.org/#1}{\includegraphics[height=0.3cm,keepaspectratio]{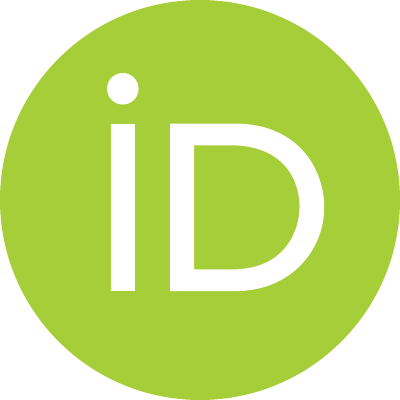}}}

\newcommand{\be}{\begin{equation}}
\newcommand{\ee}{\end{equation}}
\newcommand{\bea}{\begin{eqnarray}}
\newcommand{\eea}{\end{eqnarray}}

\newcommand{\la}{\langle}
\newcommand{\ra}{\rangle}

\newcommand{\gev}{\text{GeV}}

\newcommand{\beq}{\begin{eqnarray}}
\newcommand{\eeq}{\end{eqnarray}}
\newcommand{\bpmatrix}{\begin{pmatrix}}
\newcommand{\epmatrix}{\end{pmatrix}}
\newcommand{\ba}{\begin{array}}
\newcommand{\ea}{\end{array}}

\hypersetup{
 colorlinks=true,
 linkcolor=verdec,
 urlcolor=verdec,
 citecolor=verdec
}

\begin{document}

\title{\color{verdes} Echoes of Veltman criteria on the next-two-Higgs-doublet model}

\author{Abdesslam Arhrib\orcid{0000-0001-5619-7189}}
\email{a.arhrib@gmail.com}
\address{FST, Abdelmalek Essaadi University, B.P. 416. Tanger, Morocco}
\author{Rachid Benbrik\orcid{0000-0002-5159-0325}}
\email{r.benbrik@uca.ac.ma}
\address{Polydisciplinary Faculty, Laboratory of Fundamental and Applied Physics, Cadi Ayyad University, Sidi Bouzid, B.P. 4162, Safi, Morocco}
\author{Larbi Rahili\orcid{0000-0002-1164-1095}}
\email{rahililarbi@gmail.com/l.rahili@uiz.ac.ma}
\address{Laboratory of Theoretical and High Energy Physics (LPTHE), Faculty of Science, Ibnou Zohr University, B.P 8106, Agadir, Morocco}
\author{Souad Semlali\orcid{0000-0002-5885-7475}}
\email{s.semlali@soton.ac.uk}
\address{School of Physics and Astronomy, University of Southampton, Southampton, SO17 1BJ, United Kingdom}
\address{Particle Physics Department, Rutherford Appleton Laboratory, Chilton, Didcot, Oxon OX11 0QX, United Kingdom}
\author{Bassim Taki\orcid{0009-0009-2642-1288}}
\email{bassim.taki10@gmail.com}
\address{Laboratory of Theoretical and High Energy Physics (LPTHE), Faculty of Science, Ibnou Zohr University, B.P 8106, Agadir, Morocco}

\begin{abstract}
We investigate how the Next-Two-Higgs Doublet Model extension (N2HDM) should look if we are to address the naturalness problem using dimensional regularization. In such a model, new Higgs states are predicted, namely: three CP-even $h_{1,2,3}$, one CP-odd $A$, and a pair of charged Higgs boson $H^\pm$. Our calculations of the overall quadratic divergences have been performed with full consistency with the latest data from the Large Hadron Collider (LHC) concerning the observed 125 GeV Higgs boson, alongside precision electroweak data tests and lower mass limits on charged Higgs boson. It is shown that the quadratically divergent quantum corrections $\delta_i$ (i=1,2,3) for the three CP-even Higgs bosons are controllably small, though hidden fine-tuning might still be required. This reveals a significant impact on the model parameter space, Higgs spectrum mass and notably the singlet-doublet admixture.
\end{abstract}

\maketitle

\section{Introduction}
\label{sec:intro}
Despite its successes in describing the fundamental particles, most recently Higgs boson with 125 GeV \cite{CMS:2012qbp, ATLAS:2012yve}, the Standard Model of particles physics (SM) falls short in providing satisfactory explanations for various questions such as neutrinos mass \cite{Altarelli:2010fk}, dark matter \cite{Bertone:2018krk}, B-meson anomalies \cite{Weinberg:1979sa, LHCb:2014cxe} and many others. One particular issue is the naturalness of the Higgs mass. From an experimental point of view,  the mass of the Higgs boson (125 GeV) is in the same order as the electroweak scale. However, from a naturalness perspective, this mass is significantly larger than what would be expected at the electroweak scale. The reason behind this discrepancy lies in the substantial radiative corrections to the Higgs mass, resulting in an unnatural fine-tuning between the tree-level Higgs mass and the radiative corrections. These radiative corrections exhibit divergences and demonstrate a quadratic sensitivity to the highest scale in the theory \cite{Susskind:1978ms}. And, to face this problem, solutions propose extending the SM so that the embedded new physics (NP) can compensate for these significant corrections to the Higgs boson mass. In this regard,  Veltman \cite{Veltman:1980mj} put forward some forty years ago, in the realm of the SM, the idea that the radiative corrections to the scalar mass should either vanish or be kept at a manageable level. Theoretically, however, such approach (often referred to by VC for Veltman condition) can mitigate the impact of radiative corrections and thus can be used as an unambiguous guide for physics beyond the SM. 

In the literature, VC was the subject of many studies beyond standard model (BSM), where the SM Higgs sector is extended either by singlets \cite{Grzadkowski:2009bp, Karahan:2014ola,aali:2020tgr}, an extra doublet \cite{Ma:2001sj, Grzadkowski:2010dn,Chakraborty:2014oma, Biswas:2014uba, Chowdhury:2015yja, Darvishi:2017bhf} or a triplet fields \cite{Chabab:2015nel, Chabab:2018ert, Ouazghour:2023eqr}. In such extensions, the VC approach is treated by either: i) \underline{dimensional regularization} (DR) \cite{Einhorn:1992um} which is convenient for handling divergences but does not fundamentally address the naturalness problem and can mask the need for fine-tuning, or via ii) \underline{cut-off regularization} \cite{Oleszczuk:1994st} which introduces a cut-off parameter $\Lambda$ that limits the range of integration, thereby providing a clear physical interpretation of the regularization process. In a recent paper by Branchina et al \cite{Branchina:2022jqc} the authors \cite{Branchina:2022jqc} provided a more comprehensive understanding of the naturalness problem using the Wilsonian renormalization group (RG) approach. They found that many proposed solutions often involve hidden adjustments when examined through the Wilsonian framework. Similarly, within the same context, conservative naturalness bounds on BSM extensions have been derived using a full two-loop RGE analysis \cite{Clarke:2016jzm}. Overall, the presence of VC leads to significant effects on corresponding parameter spaces and on Higgs phenomenologies in BSM.  

In this work, we underline an extension wherein the 2HDM \cite{Gunion:1989we, Branco:2011iw} Lagrangian is extended by a real scalar singlet $S$. The resulting model, referred to as the N2HDM, has been extensively studied and is considered a potential benchmark in experimental analysis, notably with regard to dark matter (DM) purposes \cite{He:2008qm, Boucenna:2011hy, Bai:2012nv, Guo:2014bha, Drozd:2015gda}, where the singlet field is inert and constitutes a viable DM candidate. However, a non-vanishing vacuum expectation value (VEV) for the singlet field is equally important and remains fertile ground for further experimental discussions \cite{Chen:2013jvg}. Indeed, this fact is emphasized throughout the mixing between the singlet field and 2HDM doublets, leading after the spontaneous electroweak symmetry breaking (EWSB), to three neutral CP-even Higgs bosons, namely: $h_1$, $h_2$ and $h_3$, making the N2HDM even more interesting phenomenologically. As a result, the projected limits of 2HDM research are expected to be altered. Additionally, the mixing will enable the direct detection of the heavy Higgs bosons (which are mostly singlet), through, for example, $gg \to h_3 \to ZZ$. Another notable aspect of N2HDM with a non-zero VEV for the singlet field is related to the couplings with SM particles, which can be sufficiently weak, allowing light Higgs bosons to evade exclusions by Higgs searches at LEP, Tevatron, and LHC in the low mass range. Additionally, like the 2HDM, the N2HDM presents four types of Yukawa interactions with no FCNC at tree level \cite{Grossman:1994jb}, depending on which type of fermions couples to which doublet $H_{i}$. By convention, $H_{2}$ is the doublet to which all fermions couple within type I, unlike type II where such doublet couples only to up-type quark whereas the remaining fermions couple to $H_{1}$.  Type III, also called flipped models, has down-type quarks coupling to the first doublet and up-type quarks and charged leptons coupling to the second doublet. Ultimately, in type IV, or the lepton-specific models, all quarks couple to the second doublet and charged leptons to the first doublet. Here, we intend to study the VC in the context of a Type-I and Type-II  N2HDM throughout a conventional dimensional regularization, which is more consistent and particularly suitable for preserving the local gauge symmetry of the underlying Lagrangian. Moreover,  we argue that the leptonic contribution is negligible when compared to the dominant top and bottom quark ones, and hence type I and type IV are identical, as are type II and type III. This provides an opportunity for focusing on types I and II and derive the corresponding VC for the three CP-even Higgs bosons $h_i\,(i=1,2,3)$.
 
The paper is organized as follows. In the next section, we briefly review the main features of N2HDM and present the complete set of theoretical and experimental constraints that the model is subject to. Section \ref{sec:VC} is devoted to deriving the modified VC in N2HDM. Finally before the conclusion in section \ref{sec:conlusion}, we present and discuss our numerical result in section \ref{sec:results}.
%
\section{N2HDM in a nutshell}
\label{sec:n2hdm_nutshell}
In this section, we present a review of the N2HDM. We discuss the scalar potential and derive the spectrum and the parametrization of the model. We also present the Yukawa textures and discuss the natural flavor conservation of the model. Couplings of the Higgs bosons to gauge bosons are also shown and their sum rules are discussed.
\subsection{The model}
\label{subsec:model}
The scalar sector of N2HDM consists of two weak isospin doublets 
$H_{i}$ (i = 1,2), with hypercharge $Y= 1$ and a real singlet field with hypercharge $Y= 0$ which are given by
\begin{eqnarray}
H_{i} & = \left(
                   \begin{array}{c}
                    \phi_i^\pm \\
                    \frac{1}{\sqrt{2}}(v_i + \phi_i + i \chi_i) \\
                    \end{array}
                    \right)~~{\rm and}~~S = v_s + \phi_s.
\end{eqnarray}

\noindent
The most general explicit form of the Lagrangian is given by
\begin{equation}
\label{eq:LagN2HDM}
\mathcal{L} = (D_\mu H_1)^\dagger(D^\mu H_1)+(D_\mu H_2)^\dagger(D^\mu H_2)+(\partial_\mu S)^\dagger(\partial^\mu S)-V(H_1,H_2,S)
\end{equation}
where, by analogy to the 2HDM, a $\mathbb{Z}_2$ symmetry is imposed and corresponds to the invariance of the Lagrangian under a general gauge transformation, $H_1$, $H_2$ and $S$ transform as $H_1 \to H_1$, $H_2 \to -H_2$ and $S \to S$. Furthermore, a mandatory second discrete $\mathbb{Z}_2$ symmetry is also required, ensuring the singlet distinct role in the potential and corresponds to the invariance of the Lagrangian under the simultaneous transformations
\begin{equation}
\label{eq:secondZ2sym}
H_1 \to H_1, \quad H_2 \to H_2, \quad S \to -S.
\end{equation}
Hence, one can then write the most general renormalizable scalar potential for the model that respect 
$SU(2)_L\otimes U(1)_Y$ gauge symmetry as following \cite{Muhlleitner:2016mzt,Arhrib:2018qmw}:
\begin{eqnarray}
V(H_1,H_2,S) &=&
m^2_{11}\, H_1^\dagger H_1
+ m^2_{22}\, H_2^\dagger H_2 - \mu_{12}^2\, \left(H_1^\dagger H_2 + H_2^\dagger H_1\right) + \frac{1}{2}m^2_S S^2 \nonumber\\
&+& \frac{\lambda_1}{2} \left( H_1^\dagger H_1 \right)^2
+ \frac{\lambda_2}{2} \left( H_2^\dagger H_2 \right)^2 + \lambda_3\, H_1^\dagger H_1\, H_2^\dagger H_2 + \lambda_4\, H_1^\dagger H_2\, H_2^\dagger H_1\nonumber\\
&+&\frac{\lambda_5}{2} \left[\left( H_1^\dagger H_2 \right)^2+ \left( H_2^\dagger H_1 \right)^2 \right] +
\frac{\lambda_6}{8} S^4 +\frac{1}{2}[\lambda_7 H_1^\dagger H_1 +\lambda_8 H_2^\dagger H_2 ] S^2
\label{eq:Vpot}
\end{eqnarray}
where $m_{11}^2, m_{22}^2$ and $m_{S}^2$ are masse terms. In the present study, we assume that all scalar couplings $\lambda_{i}\,(i=1,2,3,4,5,6,7,8)$ are dimensionless real parameters and similarly $\mu_{12}^2$ (which softly breaks the first $\mathbb{Z}_2$ symmetry mentioned above). 

Given that, once the EWSB is taking place at some electrically neutral point in the field space, six physical Higgs states are generated: three CP-even scalars (namely $h_1$, $h_2$ and $h_3$ with $m_{h_1}< m_{h_2}< m_{h_3}$), one CP-odd scalar ($A$) and a charged Higgs pair ($H^\pm$). The squared-masses of the CP-odd and charged Higgs states do not change with respect to the 2HDM, and lead together to \cite{Gunion:2002zf},
\begin{eqnarray}
m_{H^\pm}^2 = m_{A}^2 + \frac{1}{2} v^2 (\lambda_5 - \lambda_4),
\label{eq:mAmHp}
\end{eqnarray}
with the corresponding squared-mass matrices can be diagonalized by the same rotation matrix 
\begin{equation}
\label{eq:rotation-matx}
\mathcal{R}_{\beta}=\begin{pmatrix}
c_\beta & s_\beta  \\
-s_\beta & c_\beta
\end{pmatrix}
\end{equation}
whose entries $c_\beta \equiv \cos(\beta)$ and $s_\beta \equiv \sin(\beta)$ define the ratio $t_\beta \equiv \tan\beta = s_\beta/c_\beta = v_2/v_1$.\\
On the other hand, the CP-even neutral sector is modified compared to the 2HDM and the corresponding mass matrix can be cast into a $3\times3$ one as following
\begin{equation}
\label{eq:ME}
{\mathcal{M}}_{E}^2=
\left(
\begin{array}{ccc}
\mu^2t_\beta + \lambda_1v^2 c^2_\beta\,   &  \, -\mu^2 + \frac{\lambda_L v^2 s_{2\beta}}{2}  \,  &   \,\lambda_7vv_S c_\beta \\
-\mu^2 + \frac{\lambda_L v^2 s_{2\beta}}{2}  \, &  \, \mu^2 t_\beta^{-1} + \lambda_2v^2 s^2_\beta\,   &   \, \lambda_8vv_S s_\beta\\
\lambda_7vv_S c_\beta \,   &   \, \lambda_8vv_S s_\beta \,   &   \,\lambda_6v_S^2
\end{array}
\right)
\end{equation}
with $\lambda_L=\lambda_3+\lambda_4+\lambda_5$. 
Additionally, in this context, a switch from the gauge basis $(\phi_1,\  \phi_2,\  \phi_s)$ to the mass basis $(h_1,\  h_2,\  h_3)$ is done by means of a $3\times3$ rotation matrix given by\footnote{For convenience, throughout this study, we use the short-hand notations $s_k$ and $c_k$ (with $k=1,2,3$) for the trigonometric functions $\sin(\alpha_k) \equiv s_{\alpha_k}$ and $\cos(\alpha_k) \equiv c_{\alpha_k}$, where the mixing angle $\alpha_k$ are allowed to lie between $-\pi/2$ and $\pi/2$},
\begin{equation}
\label{eq:RotMat}
\mathcal{R} = 
\left(
\begin{array}{ccc}
 c_1 c_2 & s_1 c_2  & s_2 \\
-c_1 s_2 s_3 - s_1 c_3 & c_1 c_3 - s_1 s_2 s_3 & c_2 s_3 \\
 -c_1 s_2 c_3 + s_1 s_3  & -c_1 s_3 - s_1 s_2 c_3 & c_2 c_3
\end{array}
\right),
\end{equation}
in such a way that
\begin{equation}
\label{eq:ortho}
h_i = \mathcal{R}_{ij} \phi_j \quad \big[i=1,2,3 \,\land\, j=1,2,s\big]
\quad \text{and} \quad
{\mathcal{M}}_{E}^2=\mathcal{R}_{ij} ^{T}
\left(
\begin{array}{ccc}
m_{h_1}^2   & 0  &   0 \\
0   & m_{h_3}^2  &   0 \\
0   & 0  &   m_{h_3}^2
\end{array}
\right)\mathcal{R}_{ij}.
\end{equation}
%

%
\subsection{Theoretical constraints }
\label{subsec:theo_subsec}
As with any extension BSM, the N2HDM parameter space must fulfill extensive theoretical requirements, which are summarized below:
\begin{itemize}
\item [$\circ$] Boundedness from below (BFB) of the potential \cite{Muhlleitner:2016mzt,Arhrib:2018qmw}.
\item [$\circ$] Perturbative unitarity \cite{Arhrib:2018qmw}.
\end{itemize}
For more details, we refer the reader to the Appendix, in which the explicit form of the eigenvalues at the tree level is given.
\begin{itemize}
\item [$\circ$] The vacuum structure: while looking closer at the aforementioned requirements, it is quite evident that breaking any of them is likely to indicate that the potential electroweak vacuum is not a minimum. Typically, various extrema may be held aside from the electroweak minimum depending upon the parameter values of the scalar potential, which inevitably had implications on the constraints on the aforementioned parameters. Nonetheless, by restricting vacuum to be global minimum, the value of the scalar potential at such electroweak 
minimum, $\la V \ra_{\text{EWSB}}$, reads:
\begin{equation}
\label{eq:vacuum1}
\la V \ra_{\text{EWSB}}=
\frac{1}{32} \Big[-4 v^4 c_\beta^4 \lambda_1 - 4 v^4 s_\beta^4 \lambda_2 + v^4 \big(-1 + c_{4\beta}\big) \lambda_L - v_S^4 \lambda_6 - 4 v^2 v_S^2 \big(c_\beta^2 \lambda_7 + s_\beta^2 \lambda_8\big)  \Big]
\end{equation}
and, thus, spontaneous electroweak symmetry breaking would be energetically disfavored if $\la V \ra_{\text{EWSB}}>0$, leading to the following requirement
\begin{equation}
\label{eq:vacuum2}
\lambda_1\,c_\beta^4 + \lambda_2\,s_\beta^4 + 2\, \lambda_L\,s_\beta^2\,c_\beta^2 + \frac{1}{4}\,\lambda_6\,\zeta^4 + (\lambda_7\,c_\beta^2 + \lambda_8\,s_\beta^2)\,\zeta^2 > 0,
\end{equation}
in which $\zeta$ stands for the ratio $\zeta \equiv v_S/v$.
\end{itemize}
%

\section{Veltman Conditions}
\label{sec:VC}
We address in this section how the radiative corrections to the observed 125 GeV mass either vanish or are kept under control at a manageable level within the framework of the N2HDM. Such requirements, the so-called Veltman condition (VC) \cite{Veltman:1980mj}, have been studied
over the years with the pioneering work of Stuckelberg \cite{Stueckelberg:1938hvi}. Consequently, based on both Lorentz and gauge invariant dimensional regularization method as in \cite{Chabab:2015nel}, we seek to establish relations between the coupling constants of N2HDM (mass relations) by collecting and computing the quadratic divergences and setting all of them to zero to the extent possible. 

Before getting into the nitty-gritty details of how to deal with the DR scheme, it may perhaps be useful to mention that the quadratic divergences of the Higgs self-energies could be derived in the symmetry unbroken phase, in terms of the original fields. Concretely, this approach has been the subject of many studies \cite{Osland:1992zh, Osland:1992cq, Newton:1993xc, Ait-Ouazghour:2020slc}, the results of which helped shed a little light on how canceling quadratic divergences could restrict space parameters and physical observables BSM. 

\begin{figure}[!hb]
\centerline{%
\includegraphics[width=0.3\textwidth]{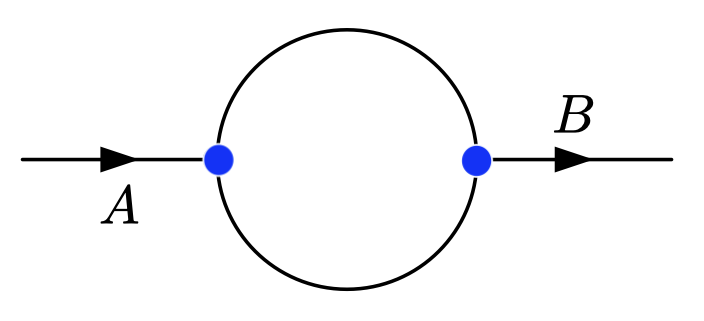}}
\caption{Higgs self-energy diagrams. $A$ and $B$ can each be 
one of the scalar fields $S$, $H_1$ or $H_2$, resulting in 9 possible diagrams.}
\label{fig-diag-1}
\end{figure}

In the N2HDM, a prior study has explored the dark doublet phase (DDP), where only one $SU(2)_{L}$ Higgs doublet and the singlet exhibit non-zero VEVs. For more details, we refer the reader to \cite{Azevedo:2021ylf}. Therefore, in our study, we consider all VEVs as non-zero, which could provide a more comprehensive understanding of the VC in the N2HDM. Thus, the relevant two-point functions representing the Higgs self-energy are displayed in Fig.~\ref{fig-diag-1}.

\begin{figure}[!h]
\centerline{%
\includegraphics[width=0.8\textwidth]{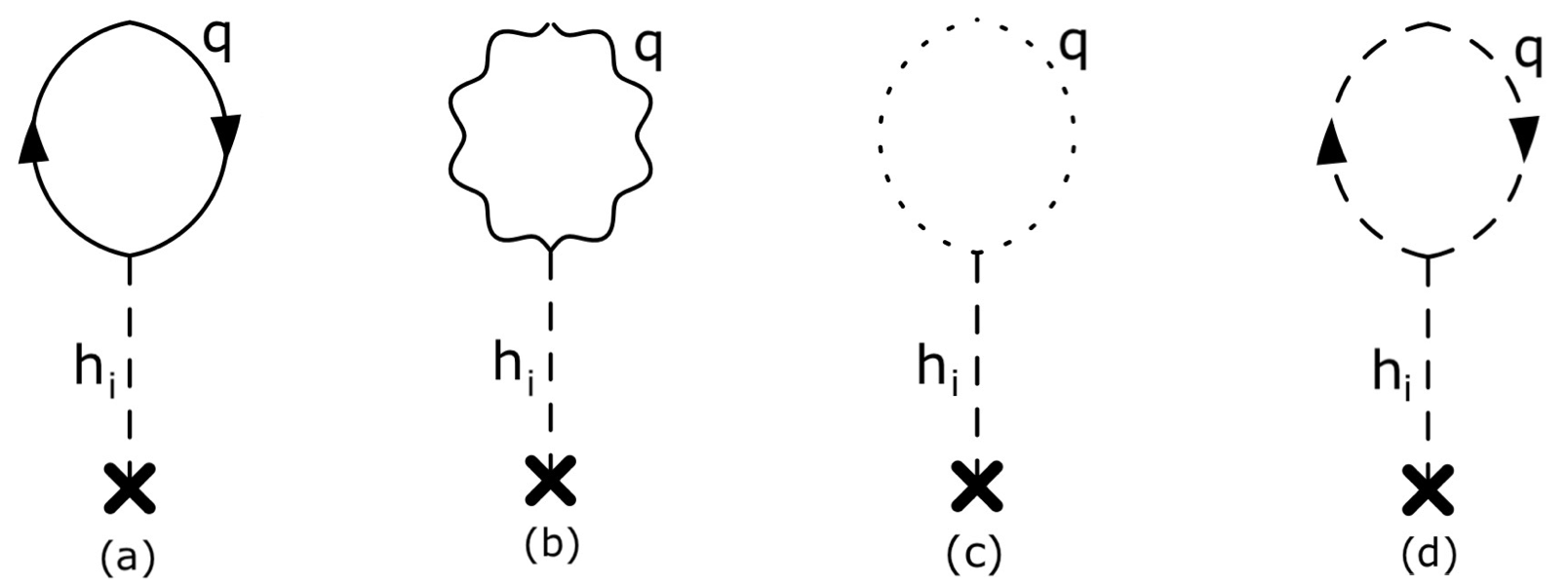}}
\caption{Higgs bosons $h_i\,(i=1,2,3)$ tadpole diagrams showing the contribution at one loop of: fermions: straight line; vector bosons: wiggly line; scalars: short dashed line and ghosts: long-dashed line.}
\label{fig-diag-2}
\end{figure}

Thereafter, in furtherance of our analysis, we will focus on the phase where the $SU(2) \times U(1)$ gauge symmetry is broken. So, by assuming that the vacuum is CP-even, one needs to calculate the quadratic divergences that show up in the tadpoles for the three CP-even neutral Higgs of our model.
Also, it is noticeable that no QCD contribution appears at one loop level, hence only the electro-weak part of the N2HDM model is concerned in this procedure. Aside from the coupling constant, we just need to consider the propagator of the field in the loop as can be seen by the topologies in Fig.~\ref{fig-diag-2},
where 
the self-energy contribution to the free propagator can be written in a simple form in terms of the Passarino-Veltman function \cite{Passarino:1978jh} as
\begin{eqnarray}
\hspace{-0.6cm} A_0(m^2) =\frac{i}{16\pi^2} \int \frac{d^nq}{q^2 - m^2} \longrightarrow \Delta_2,
\end{eqnarray}
in which $\Delta_2$ is a pure U.V. divergent number; stands for the pole term in the chosen $n=2$ dimensional space-time.

\begin{table}[!h]
\vspace*{2ex}
\centering
\renewcommand{\arraystretch}{2}
\begin{tabular}{c c c c}
\hline\hline 
Type particle &&& Leading contribution \\
\hline\hline  
scalar: $h_1, h_2, h_3, A, G, G^{\pm},\eta^Z, \eta^{\pm}$   &&&   $\frac{i}{q^2 - m^2} \times A_0(m^2)$ \\\hline
vector: $W^{\pm},Z$      &&& $-i \big(\frac{ T^{\mu \nu}}{q^2 - m^2} + \xi \frac{L^{\mu \nu}}{q^2 - \xi m^2}\big) \times \Big[(n-1) A_0(m^2) + \xi A_0(\xi m^2)\Big]$   \\ \hline 
fermion: $\it l$, $q$        &&& $i \frac{\gamma.q + m}{q^2 - m^2} \times m A_0(m^2)$ \\
\hline\hline  
\end{tabular}
\caption{the various propagator and their contributions in the N2HDM} 
\label{table1}
\end{table}

\noindent

By assuming a $R_{\xi}$ Feynman-'t Hooft gauge-invariant generalization, we summarize in Tab.~\ref{table1}, all the scalar, vectorial, and fermionic propagator contributions. The $(G, G^{\pm})$ and $(\eta^Z, \eta^{\pm})$ refer, respectively, to the Goldston bosons and Faddeev--Popov ghosts, while $T^{\mu \nu}$ and $L^{\mu \nu}$ are the transverse and the longitudinal projectors.

We begin by deriving the three independent relations necessary to eliminate the quadratic divergences
from the N2HDM model. To achieve this, we sum up all the possible diagrams, taking into account the $-1$ for the fermionic loops (including Faddeev-Popov ghosts), the symmetry factor $s_i$ of the diagram $i$, and possibly the color factor for the quarks, which we omit here for conciseness. Consequently, for each Higgs particle $h_k$ ($k=1,2,3$), one obtains:
\begin{equation}
T_{h_k} = \sum\limits_{i=1}^{9} c_i^{h_k} s_i^{h_k} t_i^{h_k} - \sum\limits_{i=U}^{D} c_{i}^{h_k} s_{i}^{h_k} t_{i}^{h_k} - \sum\limits_{i=10}^{11} c_i^{h_k} s_i^{h_k} t_i^{h_k} \label{eq:tad}
\end{equation}
where the couplings $c_i^{h_k}$, the symmetry factors $s_i^{h_k}$, and the propagator loops $t_i^{h_k}$ for each CP-even neutral Higgs boson $h_k$ are given in Appendix.~\ref{appdice:feynmanRules}. We should note here that only the dominate top and bottom quarks contributions are considered $(m_D \to m_b$, $m_U \to m_t)$, multiplied by the color factor. 

To satisfy the VCs, the quadratic divergences of the three tadpoles $T_{h_1}$, $T_{h_2}$, and $T_{h_3}$ for the $h_1$, $h_2$ and $h_3$ CP-even neutral scalar fields must vanish. However, this requirement may be reformulated differently, given that the generated expressions are rather long and complicated. Indeed, by using the inverse of the matrix $\mathcal{R}$ mentioned above, significant benefits arise from the linear combination of the fermionic coupling constants 
\begin{eqnarray}
\mathcal{R}^{-1}_{i1} c_{f \bar f}^{h_1} + \mathcal{R}^{-1}_{i2} c_{f \bar f}^{h_2} + \mathcal{R}^{-1}_{i3} c_{f \bar f}^{h_3} =0 \quad \forall\, i=1,2,3;
\end{eqnarray}
further it turns out that
\begin{eqnarray}
\mathcal{R}^{-1}_{i1} T_{h_1} + \mathcal{R}^{-1}_{i2} T_{h_2} + \mathcal{R}^{-1}_{i3} T_{h_3} =0 \quad \forall\, i=1,2,3
\end{eqnarray}
where its is straightforward to understand that all mixing angles disappear. Hence, the cancellation of the quadratic divergences at one-loop applied to N2HDM extension leads to a set of three conditions for type-I:
\begin{eqnarray}
&&\hspace{-0.5cm} (6m_W^2+3m_Z^2) + v^2(3 \lambda_1 + 2 \lambda_3 + \lambda_4+\frac{\lambda_7}{4}) = 0 \label{tad-H1-t1}\\
&&\hspace{-0.5cm} (6m_W^2+3m_Z^2) + v^2(3 \lambda_2 + 2 \lambda_3 + \lambda_4+\frac{\lambda_8}{4}) =\frac{12\big(m_t^2+m_b^2\big)}{s_\beta^2} \label{tad-H2-t1}\\
&&\hspace{-0.5cm} (\frac{3}{8} \lambda_6 + \lambda_7 + \lambda_8)\,v\,v_S =0
\label{tad-S-t1}
\end{eqnarray}
whereas in type-II, one can get:
\begin{eqnarray}
&&\hspace{-0.5cm} (6m_W^2+3m_Z^2) + v^2(3 \lambda_1 + 2 \lambda_3 + \lambda_4+\frac{\lambda_7}{4}) = \frac{12m_b^2}{c_\beta^2} \label{tad-H1-t2}\\
&&\hspace{-0.5cm} (6m_W^2+3m_Z^2) + v^2(3 \lambda_2 + 2 \lambda_3 + \lambda_4+\frac{\lambda_8}{4}) =\frac{12m_t^2}{s_\beta^2} \label{tad-H2-t2}\\
&&\hspace{-0.5cm} (\frac{3}{8} \lambda_6 + \lambda_7 + \lambda_8)\,v\,v_S =0,
\label{tad-S-t2}
\end{eqnarray}
where we have introduced the SM {\it vev}, .i.e. $v=\sqrt{v_1^2+v_2^2}=246$ GeV and the Weinberg angle $\theta_w$ such as $c_w= \cos{\theta_{w}}$. Moreover, these equations can be re-written in compact form in terms of the electroweak scale as well as the singlet {\it vev} either in type-I or type-II. For example, the type-II N2HDM one-loop condition of the quadratic divergences can be expressed through the following:
\begin{eqnarray}
&& \text{Eq}.(\ref{tad-H1-t2}) \to \frac{\delta T_1}{v^2} = \Big[-\frac{12m_b^2}{v^2 c_\beta^2} +\big(3 \lambda_1 + 2 \lambda_3 + \lambda_4+\frac{\lambda_7}{4}\big) +\frac{3m_W^2}{v^2}\big(2+\frac{1}{c_W^2}\big)  \Big] \le \epsilon \label{tad-H1-t22}\\
&& \text{Eq}.(\ref{tad-H2-t2}) \to \frac{\delta T_2}{v^2} = \Big[-\frac{12m_t^2}{v^2 s_\beta^2} +\big(3 \lambda_2 + 2 \lambda_3 + \lambda_4+\frac{\lambda_8}{4}\big) +\frac{3m_W^2}{v^2}\big(2+\frac{1}{c_W^2}\big)  \Big] \le \epsilon\label{tad-H2-t22}\\
&& \text{Eq}.(\ref{tad-S-t2}) \to \frac{\delta T_3}{vv_S} = \Big[\frac{3}{8} \lambda_6 + \lambda_7 + \lambda_8  \Big] \le \epsilon \label{tad-S-t22}
\end{eqnarray}
Thus, as far as $\delta T_1$, $\delta T_2$, and $\delta T_3$ are simultaneously close to zero, the VCs are satisfied. Usually, however, we generally assume that the ratios $\delta T_i/v^2$ (i=1,2) and $\delta T_3/v/v_S$ should not exceed an upper magnitude controlled by the dimensionless parameter  $\epsilon$.

Additionally, it is pertinent to emphasize that $\lambda_1$, $\lambda_2$, and $\lambda_6$ repeat only once, respectively in Eqs. (\ref{tad-H1-t2}), (\ref{tad-H2-t2}) and (\ref{tad-S-t2}), which is obvious since they are the coupling constants of the pure $H_{1,2}$ doublets and $S$ singlet quartic interactions.
Moreover and perhaps most importantly, the VCs in the 2HDM type-II given in Ref.\cite{Darvishi:2017bhf} are easily recovered, by setting to zero, the couplings $\lambda_6$, $\lambda_7$ and $\lambda_8$ in the previous formula $T_i$, so discarding any mixing between doublets and singlet.  Also, if the Yukawa couplings are neglected, the VCs for $H_1$ and $H_2$ are the same for all N2HDM, which confirms the results of Ref \cite{Chakraborty:2014oma}.

Last and not least, since all graphs contributing to the quadratic divergences of the Higgs tadpoles $T_{h_1}$, $T_{h_2}$, and $T_{h_3}$ have been evaluated, the considerations of this section lead to expressing the $\lambda$'s couplings in termes of physical masses and the $\mu_{12}^2$ parameter. For the type-II 2HDM, where all our subsequent calculations have been done, one can read,
\begin{eqnarray}
\left\lbrace\begin{matrix}
(\ref{tad-H1-t2}) \\
(\ref{tad-H2-t2}) \\
(\ref{tad-S-t2})
\end{matrix}\right.
\Leftrightarrow  
\begin{bmatrix}
\delta_1  \\
\delta_2 \\
\delta_3
\end{bmatrix}=
\begin{bmatrix}
A_{11} & A_{12} & A_{13}  \\
A_{21} & A_{22} & A_{23}  \\
A_{31} & A_{32} & A_{33}  
\end{bmatrix}\times
\begin{bmatrix}
m_{h_1}^2  \\
m_{h_2}^2  \\
m_{h_3}^2  \\
\end{bmatrix}
\label{new-tad-type2}
\end{eqnarray}
where the square mass dimension $\delta_i$ (i=1,2,3) parameters that be made
zero, or at least controllably small, by some symmetry, are given by
\begin{eqnarray}
\delta_1 & = & \frac{12 m_b^2}{\cos\beta^2} + \frac{\mu_{12}^2}{\sin\beta \cos\beta}(1+3\tan\beta^2) - m_A^2 - 2 m_{H^\pm}^2 - 6 m_W^2 - 3 m_Z^2 \label{eq:delta1}\\
\delta_2 & = & \frac{12 m_t^2}{\sin\beta^2} + \frac{\mu_{12}^2}{\sin\beta \cos\beta}(1+3\cot\beta^2) - m_A^2 - 2 m_{H^\pm}^2 - 6 m_W^2 - 3 m_Z^2 \label{eq:delta2}\\
\delta_3 & = & 0 \label{eq:delta3}
\end{eqnarray}
and the matrix elements $A_{ij}$ read as follows
\begin{eqnarray}
A_{1i} & = & \frac{ \mathcal{R}_{i1} \sec\beta \Big[v \mathcal{R}_{i3}+8 v_S \csc\beta \mathcal{R}_{i2} + 12 v_S \sec\beta \mathcal{R}_{i1}\Big] }{4 v_S} \label{eq:A1i}\nonumber\\
A_{2i} & = & \frac{ \mathcal{R}_{i2} \csc\beta \Big[v \mathcal{R}_{i3}+12 v_S \csc\beta \mathcal{R}_{i2} + 8 v_S \sec\beta \mathcal{R}_{i1}\Big] }{4 v_S} \label{eq:A2i}\nonumber\\
A_{3i} & = & \frac{ \mathcal{R}_{i3} \Big[3 v \mathcal{R}_{i3}+8 v_S \csc\beta \mathcal{R}_{i2} + 8 v_S \sec\beta \mathcal{R}_{i1}\Big] }{8 v_S} \label{eq:A3i}
\end{eqnarray}
with the $\mathcal{R}_{ij}$ are the matrix elements in Eq.~(\ref{eq:RotMat}) while $\sec(x)\,,\csc(x)$ are the trigonometric functions defined respectively by the inverse of $\sin(x)\,,\cos(x)$.

\section{Results}
\label{sec:results}
In the following, we present our findings assuming that $h_3$ is almost singlet, while tacitly distinguishing between two SM-like limits: (i) the SM-like $h_1$ scenario, in which the neutral Higgs partners, $h_2$ and $h_3$, are heavier (with $m_{h_2}<m_{h_3}$), and (ii) the $h_2$ SM-like limit, an intermediate state between two Higgs partners, with $h_1$ being light and $h_3$ heavier. 
\begin{table}[!h]
\begin{center}
\begin{tabular}{c|ccc}
\toprule
 & $\chi_t^{h_i}$ & $\chi_d^{h_i}$ & $\chi_V^{h_i}$ \\ \midrule
$H_1$ & $(c_{\alpha_2} s_{\alpha_1} )/s_\beta$ & $(c_{\alpha_1}
c_{\alpha_2}) / c_\beta$ & $c_{\beta-\alpha_1} c_{\alpha_2}$        \\
$H_2$ & $(c_{\alpha_1} c_{\alpha_3} - s_{\alpha_1} s_{\alpha_2}
s_{\alpha_3})/s_\beta$ & $-(c_{\alpha_3} s_{\alpha_1}+ c_{\alpha_1}
s_{\alpha_2} s_{\alpha_3})/c_\beta$ & 
$-c_{\beta-\alpha_1} s_{\alpha_2} s_{\alpha_3}+c_{\alpha_3} s_{\beta-\alpha_1}$ \\
$H_3$ & $-(c_{\alpha_1} s_{\alpha_3} + c_{\alpha_3} s_{\alpha_1}
s_{\alpha_2} )/s_\beta$ & $(s_{\alpha_1} s_{\alpha_3} - c_{\alpha_1}
c_{\alpha_3} s_{\alpha_2} )/c_\beta$ & $-c_{\beta-\alpha_1} s_{\alpha_2} c_{\alpha_3}-s_{\alpha_3} s_{\beta-\alpha_1}$ \\ \bottomrule
\end{tabular}
\caption{The modifier couplings $\chi_t^{h_i}$, $\chi_d^{h_i}$ and $\chi_V^{h_i}$ of the N2HDM Higgs bosons $h_i$, in type II.}
\label{Type2-Hiff-HiVV}
\end{center}
\end{table}

Before diving into the search for solutions to VC and the corresponding analysis, we would like to point out that in the rest of this study, we confine ourselves to the type-II N2HDM, however, the same methodology remains valid for type-I. In this regard, owing to their importance in the next calculations, we provide in Tab.~\ref{Type2-Hiff-HiVV} all the CP-even Higgs boson modifier couplings to quarks ($\chi_t^{h_i}$, $\chi_b^{h_i}$) and gauge bosons ($\chi_V^{h_i}$), expressed in terms of the fields admixture of the mass eigenstates $h_i$ and the mixing angle $\beta$. We recall here that the 2HDM limit for $h_{1,2}$ couplings to the gauge bosons \cite{Muhlleitner:2016mzt,Arhrib:2018qmw}, in both scenarios, may be achieved by setting
\begin{alignat}{2}
&h_1-{\rm scenario}: &\quad &  \beta-\alpha_1=0 \quad {\rm and} \quad \alpha_2=0,\label{scenario1}\\
&h_2-{\rm scenario}: &\quad &  \beta-\alpha_1=\frac{\pi}{2} \quad {\rm and} \quad \alpha_3=0,\label{scenario2}
\end{alignat}

\subsection{Approximate solution for VCs}
\label{subsec:approx-sol}
To analytically handle the main purpose within N2HDM type-II, we should look for a region where the Eqs.~(\ref{tad-H1-t2}), (\ref{tad-H2-t2}) and (\ref{tad-S-t2}) are released automatically. Towards this end, we note that the difference of $\lambda_1$ and $\lambda_2$ can be extracted by two alternative methods, either by using the first two diagonal elements in ${\mathcal{M}}_{E}^2$ using both Eqs. (\ref{eq:ME}) and (\ref{eq:ortho}) that express $\lambda_1-\lambda_2$ in terms of $\tan\beta$, $\mu_{12}^2$ and the mass of the neutral partners depending on the scenario, or by evaluating: (\ref{tad-H1-t2})-(\ref{tad-H2-t2}). In the latest case, the $\lambda_7$ and $\lambda_8$ couplings must be written in terms of the parameters of the physical basis as
\begin{eqnarray}
\lambda_6 &=& \frac{1}{v_S^2} \sum_{i=1}^3 m_{h_i}^2 \mathcal{R}_{i3}^2 \label{eq:la6}\\
\lambda_7 &=& \frac{1}{v v_S \cos\beta} \sum_{i=1}^3 m_{h_i}^2 \mathcal{R}_{i1}
\mathcal{R}_{i3} \label{eq:la7}\\
\lambda_8 &=& \frac{1}{v v_S \sin\beta} \sum_{i=1}^3 m_{h_i}^2 \mathcal{R}_{i2}
\mathcal{R}_{i3} \;. \label{eq:la8}
\end{eqnarray}

\noindent
Combining all the above together, and assuming both limits mentioned in Eqs.(\ref{scenario1}) (\ref{scenario2}), we obtain the following expressions for partner masses squared of the SM-like $h_1$ or $h_2$ particle as,
\begin{alignat}{2}
&\underline{h_1-{\rm scenario}}: \nonumber\\
& m_{h_2}^2=\frac{384 v_S c_3 \Big[ 2 m_t^2 - 2 m_b^2 t_\beta^2 + \mu_{12}^2 (1-t_\beta^4)/(2 t_\beta) \Big]\Big[ v_S (1-t_\beta^2) s_3 - v t_\beta c_3 \Big]}{ s_{23} \Big[ (v^2+48v_S^2)((1-t_\beta^2)^2-4 t_\beta^2)- (v^2-48v_S^2)(1+t_\beta^2)^2) \Big] - 192 v v_S c_{23} t_\beta (1-t_\beta^2)}\label{mh22-scenario1}\\
& m_{h_3}^2=\frac{384 v_S c_3 \Big[ 2 m_t^2 - 2 m_b^2 t_\beta^2 + \mu_{12}^2 (1-t_\beta^4)/(2 t_\beta) \Big] s_3 \Big[ v_S (1-t_\beta^2) - v t_\beta t_3 \Big]}{ s_{23} \Big[ (v^2+48v_S^2)((1-t_\beta^2)^2-4 t_\beta^2)- (v^2-48v_S^2)(1+t_\beta^2)^2) \Big] - 192 v v_S c_{23} t_\beta (1-t_\beta^2)}\label{mh32-scenario1}
\end{alignat}
with $m_{h_2}^2=m_{h_2}^2(t_\beta,\alpha_3,\mu_{12}^2)$ and $m_{h_3}^2=m_{h_3}^2(t_\beta,\alpha_3,\mu_{12}^2)$

\begin{alignat}{2}
&\underline{h_2-{\rm scenario}}: \nonumber\\
& m_{h_1}^2=\frac{192 v_S \Big[ 2 m_t^2 - 2 m_b^2 t_\beta^2 + \mu_{12}^2 (1-t_\beta^4)/(2 t_\beta) \Big]\Big[ v_S (1-t_\beta^2) s_{22} + 2 v t_\beta c_2^2 \Big]}{ s_{22} \Big[ (v^2+48v_S^2)((1-t_\beta^2)^2-4 t_\beta^2)- (v^2-48v_S^2)(1+t_\beta^2)^2) \Big] + 192 v v_S c_{22} t_\beta (1-t_\beta^2)}\label{mh12-scenario2}\\
& m_{h_3}^2=\frac{192 v_S \Big[ 2 m_t^2 - 2 m_b^2 t_\beta^2 + \mu_{12}^2 (1-t_\beta^4)/(2 t_\beta) \Big]\Big[ v_S (1-t_\beta^2) s_{22} - 2 v t_\beta s_2^2 \Big]}{ s_{22} \Big[ (v^2+48v_S^2)((1-t_\beta^2)^2-4 t_\beta^2)- (v^2-48v_S^2)(1+t_\beta^2)^2) \Big] + 192 v v_S c_{22} t_\beta (1-t_\beta^2)}\label{mh32-scenario2}
\end{alignat}
with $m_{h_1}^2=m_{h_1}^2(t_\beta,\alpha_2,\mu_{12}^2)$ and $m_{h_3}^2=m_{h_3}^2(t_\beta,\alpha_2,\mu_{12}^2)$, whereas $s_{2i}$ and $c_{2i}$ stand, respectively, for $\cos2\alpha_i$ and $\sin2\alpha_i$ respectively.

An interesting fact is that these masses, for each scenario, do not depend on the $m_{125}$, and the limit $\alpha_3\to0$ (resp. $\alpha_2\to0$) means that very small masses are analytically devoted to $m_{h_3}$ for $h_1$-scenario (resp. $h_2$-scenario). To better illustrate this and other related points, we have plotted, regardless of any theoretical or experimental constraints, the Higgs partner masses versus $\tan\beta$, as given by Eqs.~(\ref{mh22-scenario1}), (\ref{mh32-scenario1}), (\ref{mh12-scenario2}) and (\ref{mh32-scenario2}), for various values of $\mu_{12}^2$ and specific $\alpha_{2,3}$ values. Hence, by drawing a clear line at 125.09 (GeV), we are referring in fact to two completely separate areas. The first one corresponds to $h_1-$scenario, in which the allowed domain for $m_{h_2}, m_{h_3}$ should be located above this line. Conversely, in the $h_2-$scenario, only $m_{h_3}$ is permitted to be above the line, whereas the ${h_1}$ Higgs boson, supposed to be the lightest, must be below in the grey area.  
\begin{figure}[!ht]
\centering
\includegraphics[width=0.32\textwidth]{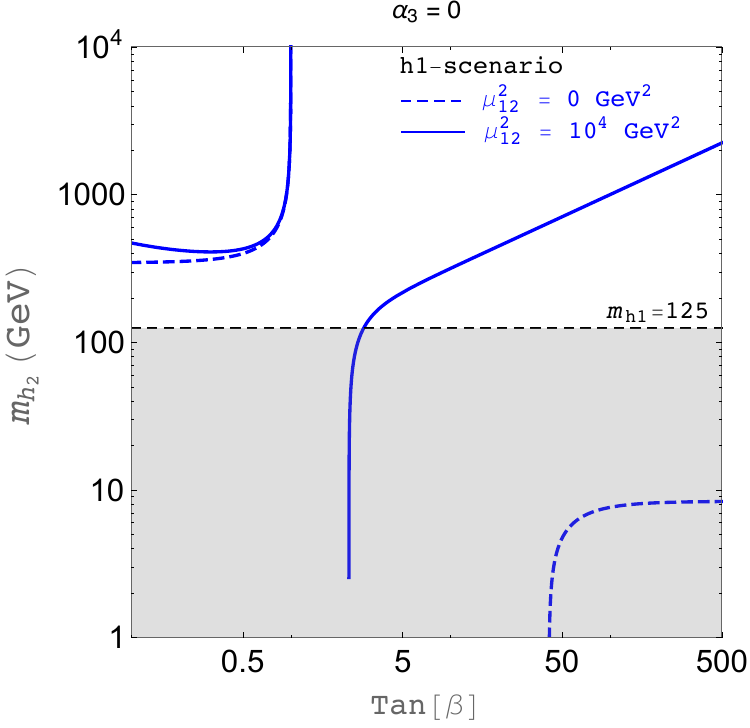} 
\includegraphics[width=0.32\textwidth]{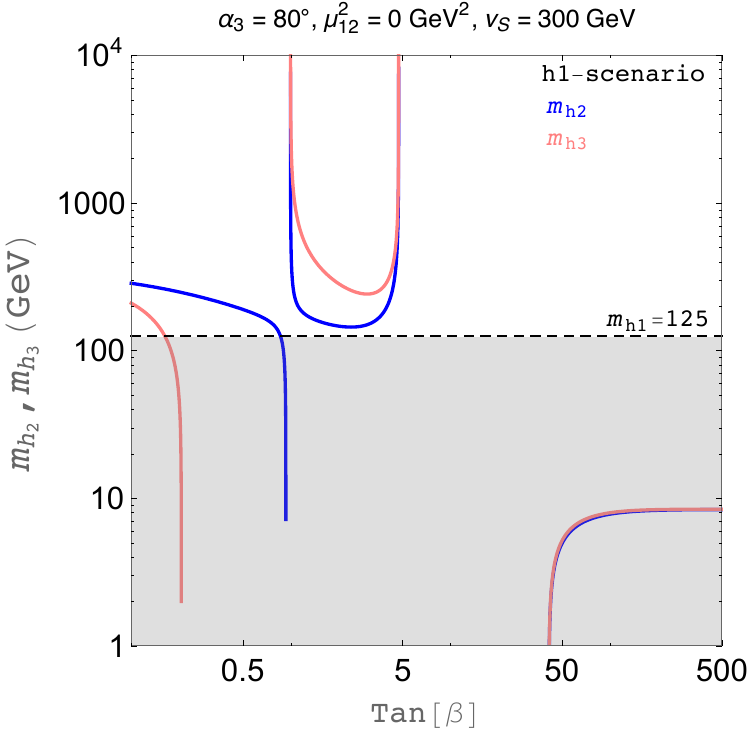} 
\includegraphics[width=0.32\textwidth]{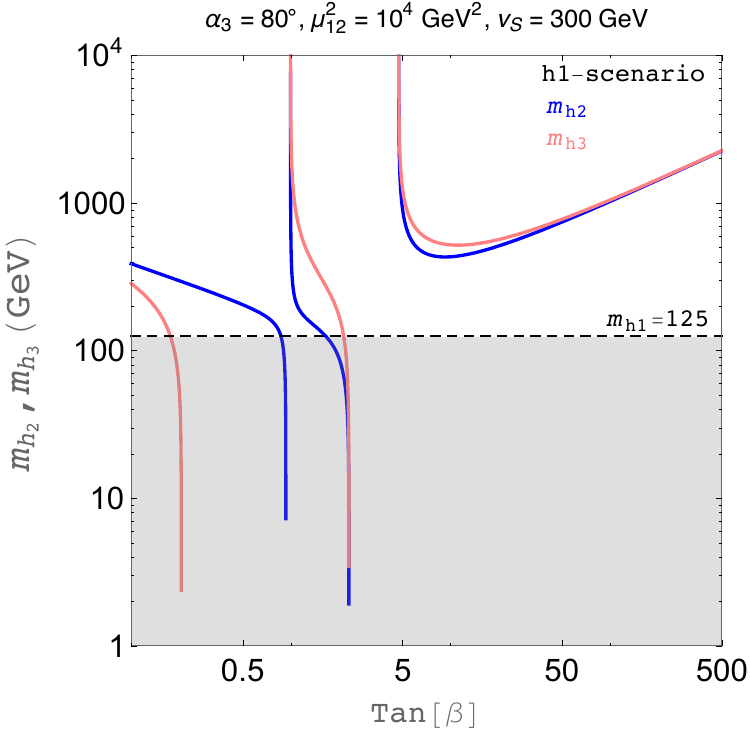}
\includegraphics[width=0.32\textwidth]{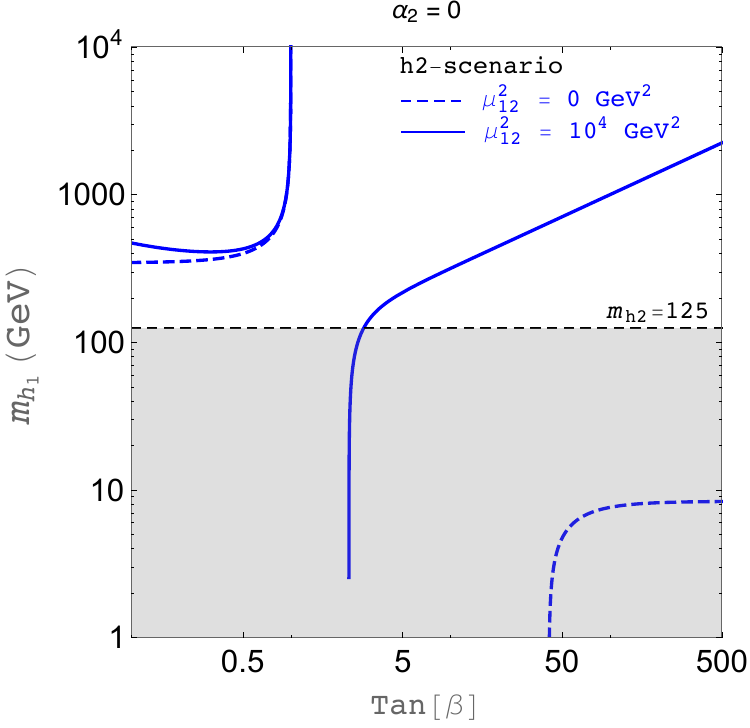}
\includegraphics[width=0.32\textwidth]{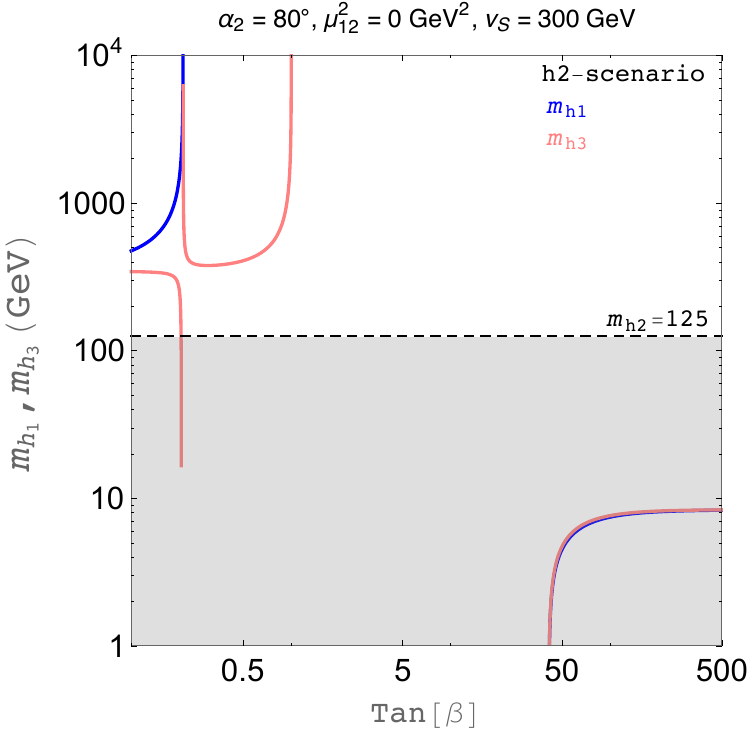}
\includegraphics[width=0.32\textwidth]{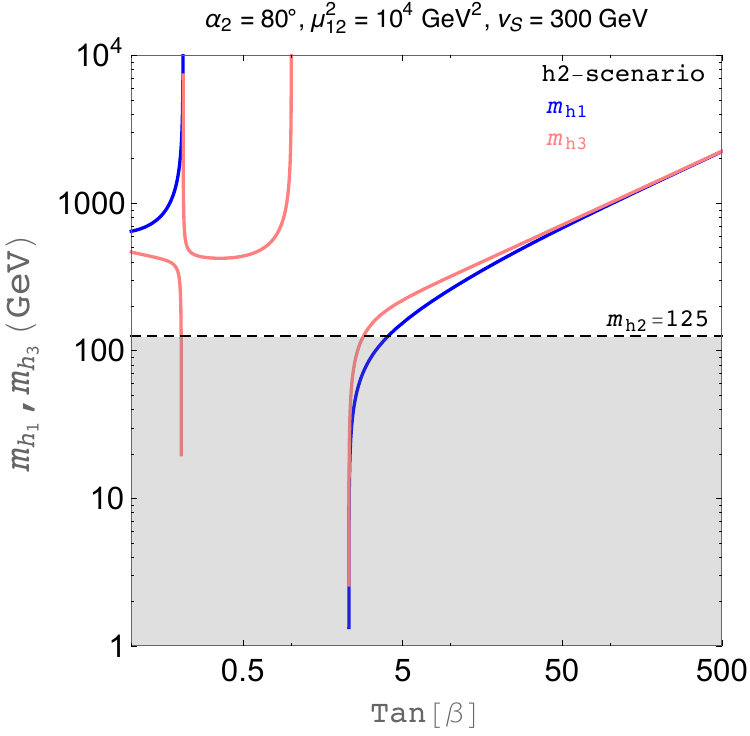}
\caption{
The mass of the partner of the SM-like $h_1(h_2)$ Higgs particle versus $\tan\beta$ based on Eq.(\ref{new-tad-type2}) for two value of $\mu_{12}^2$ and fixed value of $\alpha_2\,(\alpha_3)$ within the N2HDM type-II. 2HDM-type II limit Ref.\cite{Darvishi:2017bhf} is achieved in the left panel.}
\label{fig:Hpartner-scenario}
\end{figure}

The first remark to note is that the 2HDM-type II limit \cite{Darvishi:2017bhf} is matched as one can see in the left side of Fig.~\ref{fig:Hpartner-scenario}, whether for the $h_1$-{\rm scenario}, or for the $h_2$-{\rm scenario}. For example, in the case of exact $\mathbb{Z}_2$ symmetry, i.e. $\mu_{12}^2=0$ GeV$^2$, the partner $h_2$ can survey only for low $\tan\beta$ with a mass above the $t\bar{t}$ threshold in the $h_1-$scenario. While for the $h_2-$scenario, the $h_1$ shall take place only for $\tan\beta \ge m_t/m_b \approx 43$, measuring about $\approx 2 m_b$ in mass and less. Once $\mu_{12}^2$ has non-zero value, the intermediate $\tan\beta$ open up. So, for $h_1-$scenario, the Higgs partner mass is large for low $\tan\beta \le 1$, e.g $\gtrsim 2 m_t$ and is typically spreading for high $\tan\beta$, while for the $h_2-$scenario there is almost no solution for the $h_1$ Higgs partner mass except for a unique value of $\tan\beta \approx 2.23$.

The picture is slightly more complicated when the singlet contribution occurs. Specifically, in the $h_1$-{\rm scenario} with $\mu_{12}^2=0$ GeV$^2$, a straight and narrow solution for the $h_2$ and $h_3$ Higgs partner masses may take place for $1 \le \tan\beta \le 4.5$, exhibiting an average splitting of 220 GeV. Whilst for $\mu_{12}^2=10^4$ GeV$^2$, this range is subsequently reduced by half, but solutions for such a scenario remain possible for high values of $\tan\beta$. Conversely, in the opposite scenario, no solution exists for $\tan\beta \gtrsim 1$ regardless of $\mu_{12}^2$ value. However, for large and positive $\mu_{12}^2$, possible non-zero masses for the Higgs partner $h_1$ and $h_3$ are intended to be all between the curves below the dashed line in the grey area.

\subsection{Experimental constraints}
\label{subsec:exp-constraints}
In addition, and for the N2HDM to be consistent with data, several experimental constraints must be met, in particular,  
\begin{itemize}
\item [$\circ$] Electroweak precision test observables (EWPT): related to the so-called oblique parameters, $S$, $T$ and $U$ \cite{Peskin:1990zt, Peskin:1991sw}, and may provide a strong indirect probe of NP BSM. These parameters that quantify deviations from the SM in terms of radiative corrections to the $W$, $Z$, and the photon self energies, receive new contributions in the framework of N2HDM resulting from both neutral $h_i$, $A$, and charged Higgs states. In this study we perform the $\chi^2$ test over 
the allowed parameter space of N2HDM taking into account the correlation between $S$ and $T$ \cite{Grimus:2007if, Grimus:2008nb, Gunion:2002zf}. Our $\chi^2_{S,T}$ is defined as:
\begin{equation}
\chi^2_{S,T} = \frac{1}{\hat{\sigma}^2_{1}(1-\rho^2)}(S - \hat{S})^2
+ \frac{1}{\hat{\sigma}^2_{2}(1-\rho^2)}(T - \hat{T})^2 -\frac{2\rho}{\hat{\sigma}_{1}\hat{\sigma}_{2}(1-\rho^2)}(S - \hat{S})(T - \hat{T}) \,, \label{eq:chi2STU}
\end{equation} 
where $\hat{S}$ and $\hat{T}$ are the measured values of $S$ and $T$, $\hat{\sigma}_{1,2}$  are their one-sigma errors and $\rho$  their correlation. Fixing $U=0$, they read ~\cite{ParticleDataGroup:2022pth}, 
\begin{equation}
S = 0.00 \pm 0.07, \quad T = 0.05\pm 0.06, \quad \rho_{S,T} = 0.92
\end{equation}
\item [$\circ$] Higgs data and direct collider searches: for the sake of evaluating the SM-like Higgs boson, we use the updated {\tt HiggsTools} library to appropriately test Higgs searches and check the Higgs signal rate constraints in the N2HDM taking into account various LEP, Tevatron and recent LHC 13 TeV search results. So, by scanning over the appropriate model parameters, we ensure that either the mass or the coupling to gauge bosons consistently falls within the experimental boundaries.
\end{itemize}

Additionally, by focusing on scenario where $\tan\beta$ is not so large (.i.e $\lesssim 12$), the mass of the charged Higgs boson is restricted to be at least approximately 580 GeV \cite{Hammad:2022wpq}, and we verify whether the obtained solutions align with the constraints imposed by the EWPO oblique parameters $S$, $T$, and $U$ taking into account the extra Higgs boson contributions in the N2HDM.

\noindent
Another important point to note here is that there are no solutions emerged when considering the $h_2$-scenario. So, throughout the rest of our discussion, we consider only the $h_1$-scenario and conduct a random scan of the N2HDM parameters, adhering to the ranges below: 
\begin{eqnarray}
\label{eq:inputspara}
&& m_{h_1} = 125\,\gev,\quad  m_{h_{2,3}} \in [130,1000]\,\gev,\quad m_{H^\pm} \in [580,\,1000]\,\gev, \nonumber\\
&& m_{A} \in [200,\,1000]\,\gev,\quad  \mu_{12}^2\in [0,\,10^6]\,\gev^2,\quad v_S \in [10^2,\,10^3]\,\,\gev   \nonumber\\
&& \frac{-\pi}{2} \le \alpha_{1,2,3} \le \frac{\pi}{2}\,,\quad  \text{and} \quad   0.5 \le \tan\beta \le 12,
\end{eqnarray}
and setting $m_t=172.44$ GeV, $m_b=4.18$ GeV, $m_W=80.38$ GeV, as well as $m_Z=91.18$ GeV.
The analytic solutions for Eq.(\ref{new-tad-type2}), derived through $\mathtt{Mathematica}$, are independently implemented into a $\mathtt{Fortran}$ program during the scan. The latter is done in a succession of cuts as follows
\begin{itemize}
\item [$\circ$] Cut1. We begin by performing a random scan of the parameters, applying the positivity, unitarity, and vacuum constraints from section \ref{subsec:theo_subsec}. We impose $m_{h_2}<m_{h_3}$ to avoid the degenerate scenario between the heavier CP-even states. 
\item [$\circ$] Cut2. In this step, the $\chi^2_{S,T} $ in Eq.(\ref{eq:chi2STU}) is considered at $2\sigma$ confidence level, and {\tt HiggsTools} is applied to ensure that the remaining points align with the requirements from the Higgs boson signal strengths and heavy Higgs searches.
\item [$\circ$] Cut3. Finally, we keep only the points in the parameter space that fits within the definition of VC (as the parameter $\epsilon$ must be controllably small, we consider the cases $\epsilon=4,6,10$).
\end{itemize}

\subsection{Scanning results}
\label{subsec:scanning-res}
Now, after scrutinizing the N2HDM space parameter by applying the BFB and the perturbative unitarity constraints and respecting the experimental ones, we investigate to what extent the Higgs spectrum of N2HDM, 
\begin{figure}[!ht]
\centering
\includegraphics[width=0.343\textwidth]{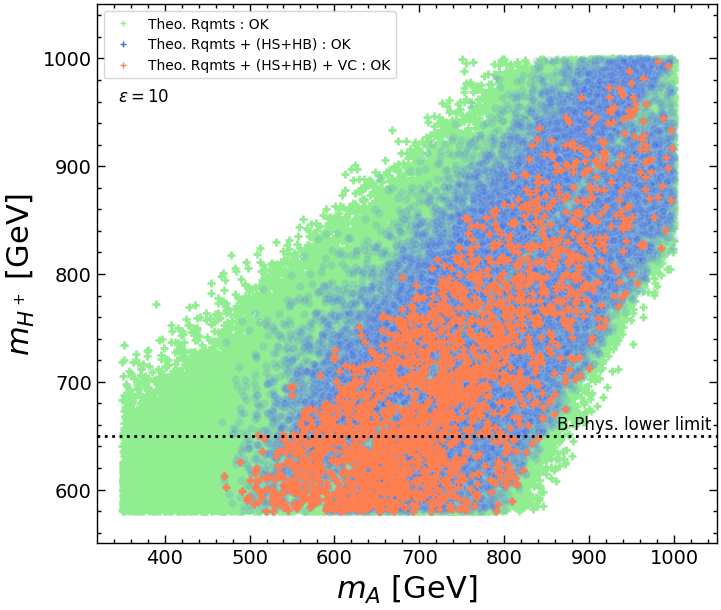} 
\includegraphics[width=0.3\textwidth]{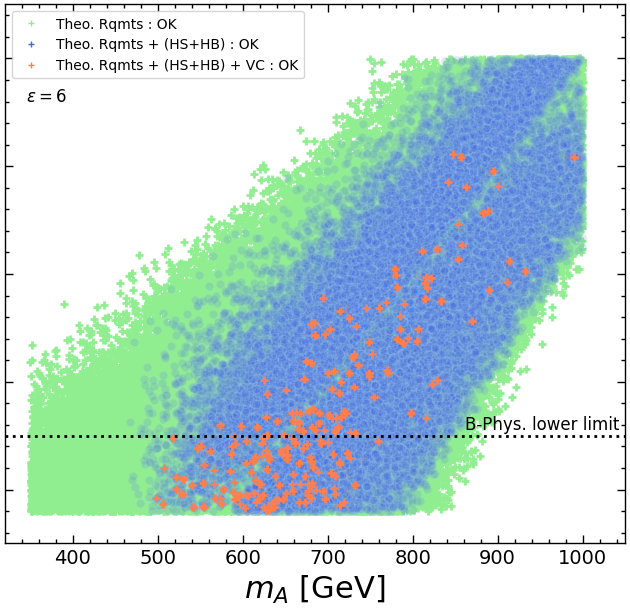} 
\includegraphics[width=0.3\textwidth]{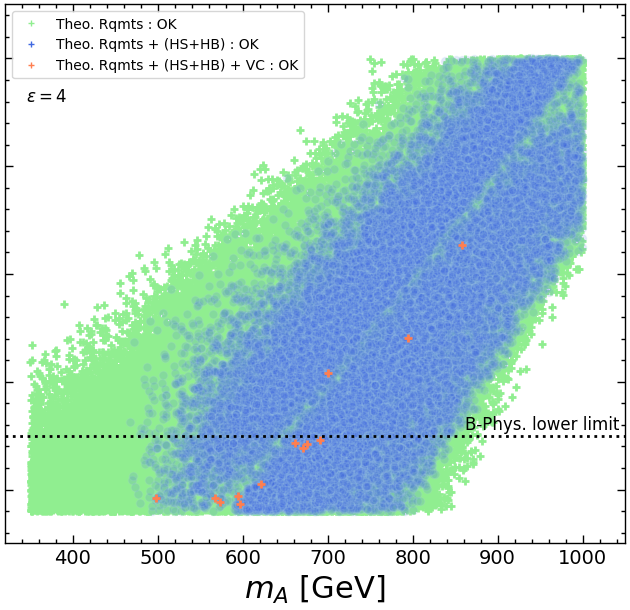} \\
\includegraphics[width=0.338\textwidth]{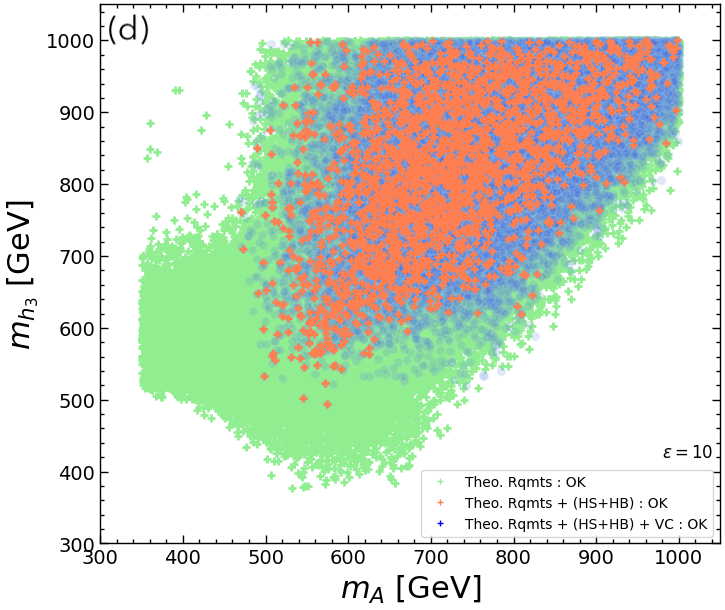} 
\includegraphics[width=0.3\textwidth]{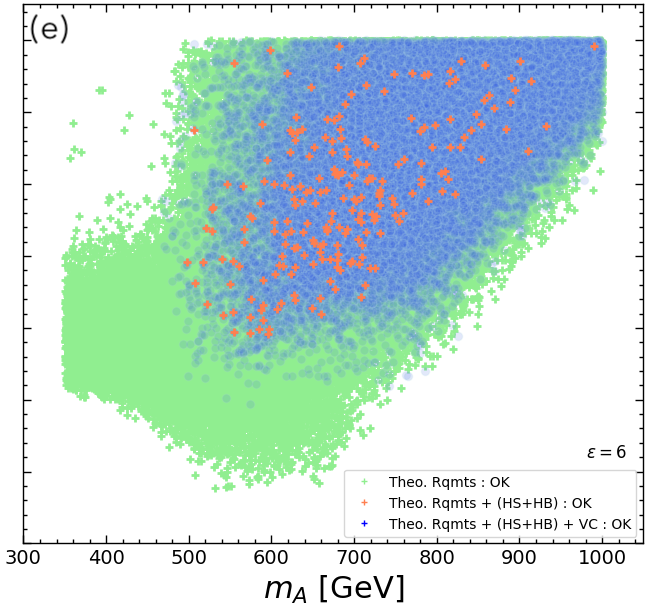} 
\includegraphics[width=0.3\textwidth]{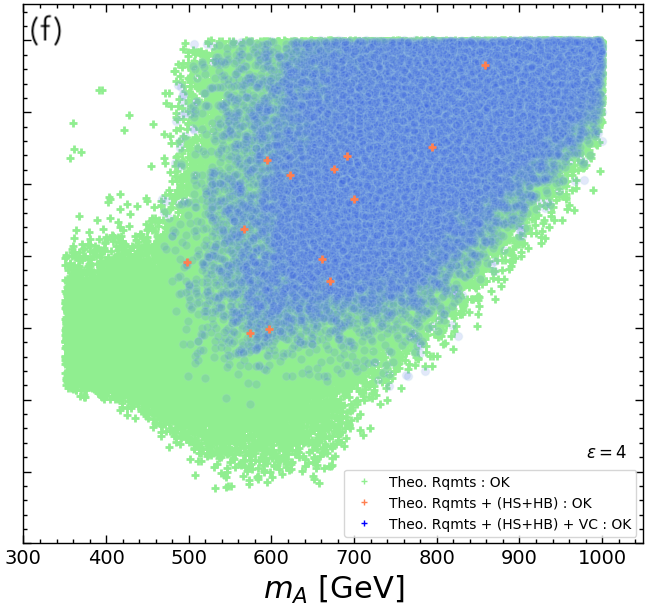} \\
\hspace{0.2cm}\includegraphics[width=0.34\textwidth]{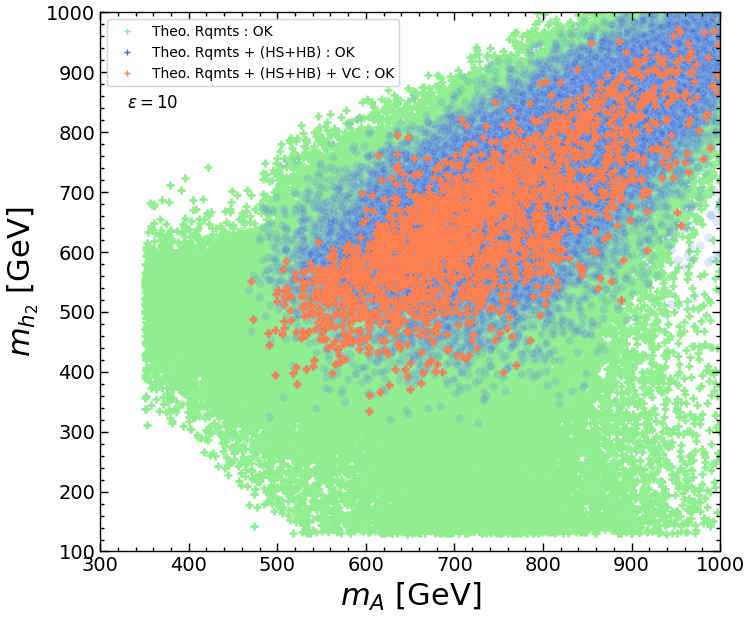} 
\includegraphics[width=0.305\textwidth]{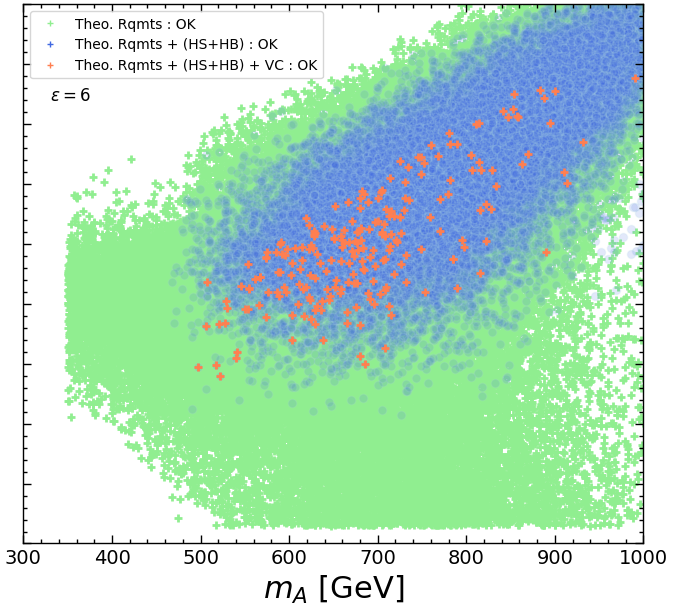} 
\includegraphics[width=0.305\textwidth]{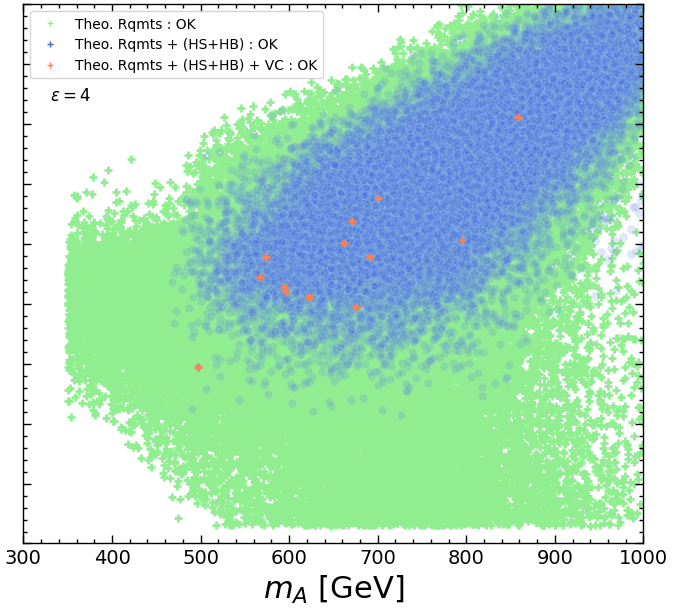} 
\caption{Outcome of the N2HDM parameter scan, incorporating theoretical constraints, bounds from Higgs signal strengths, LHC searches for additional Higgs states, and satisfaction of the Veltman conditions. Points satisfying these constraints are displayed in the planes: $[m_{A},\,m_{H^\pm}]$ (upper side), $[m_{A},\,m_{h_3}]$ (middle side), and $[m_{A},\,m_{h_2}]$ (lower side). The parameter $\epsilon$ is set to $10$, $6$, and $4$ from left to right. The horizontal dotted line represents the lower limit for the $H^\pm$ mass ($m_{H^\pm} \gtrsim 650$ GeV), derived from flavor-physics observables as outlined in Ref.~\cite{Haller:2018nnx}.}
\label{fig:masses-correlation}
\end{figure}
specifically $h_2$, $h_3$, $A$ and $H^\pm$, could be probed through such constraints. In addition; by providing our computational analysis of Eq.(\ref{new-tad-type2}), we specifically highlight naturalness to demonstrate the VC effect on those Higgs masses. 

In Fig.~\ref{fig:masses-correlation}, we exhibit the correlation between the CP-odd Higgs boson mass, $m_A$ and the other scalars masses, namely: $m_{h_2}$, $m_{h_3}$ and $m_{H^\pm}$. We present the parameter space regions allowed by theoretical requirements in green, and those further validated by  {\tt Higgsbounds} and  {\tt Higgssignals} in blue. The orange samples reflect the remaining points after considering the Veltman conditions for three values of the $\epsilon$ parameter. The dotted black horizontal line in the upper panel indicates the lower limit for the $H^\pm$ mass; i.e. $m_{H^\pm} \gtrsim 650$ GeV, derived from flavor-physics observables \cite{Haller:2018nnx}. It is obvious that by decreasing the $\epsilon$ parameter, the charged Higgs mass as well as $m_A$ get drastically restricted - for instance, for $\epsilon=4$ and taking into account limit from B-physics - most of the orange points falling within the range of $m_A$  (resp. $m_{H^\pm}$), from 700 GeV to 858 GeV (resp. from 706 GeV to 824 GeV), as can be seen from upper left panel in Fig.~\ref{fig:masses-correlation}.

The other two CP-even non-SM-like neutral Higgs bosons have similar situations and are relatively heavy and strongly constrained by naturalness as can be seen from the middle and lower panels in Fig.~\ref{fig:masses-correlation}. For $h_2$ (resp. $h_3$), the excluded Higgs mass region is significantly extended with lower bounds around 604 GeV (resp. 776 GeV) and upper bounds reaching 809 GeV (resp. 962 GeV).
\begin{figure}[!h]
\centering
\includegraphics[width=0.348\textwidth]{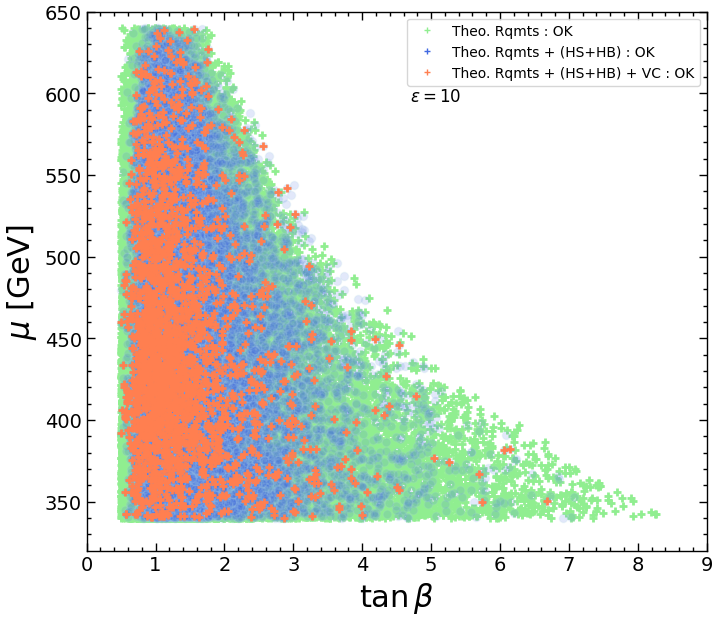} 
\includegraphics[width=0.3115\textwidth]{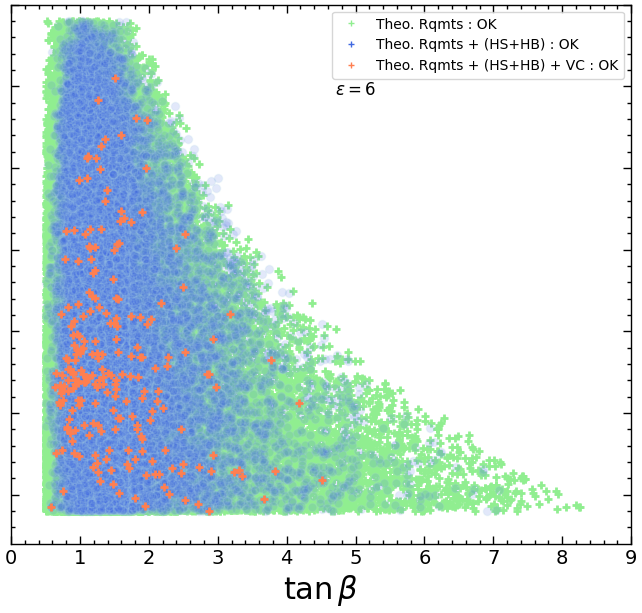} 
\includegraphics[width=0.3115\textwidth]{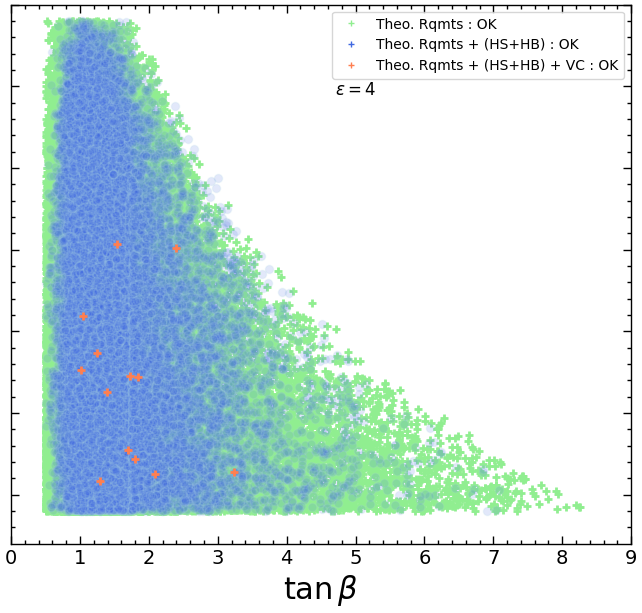}
\caption{The allowed regions in the $[\tan\beta, \mu_{12}]$ plane are depicted after considering the VC. The color coding and inputs are the same as in Figure \ref{fig:masses-correlation}.}
\label{fig:mu12-tb-correlation}
\end{figure}

Moreover, a similar analysis is also performed in the $[\tan\beta, \mu_{12}]$ plane, and Fig.~\ref{fig:mu12-tb-correlation} makes it clear that if naturalness induced conditions are imposed, the allowed parameter space is reduced even more and only a few points remain viable, represented in orange, as the value of $\epsilon$ parameter decreases. More precisely, we find that $\mu_{12}$ and $\tan\beta$ parameters are particularly sensitive to the naturalness conditions, mainly to $\delta{m_{h_1}}$. As a result, solutions for Eq.(\ref{new-tad-type2}) may only occur for $\tan\beta$ slightly below $3.2$, and for positive $\mu_{12}$, in the range $200\sim510$ GeV. 

\begin{figure}[!h]
\centering
\includegraphics[width=0.328\textwidth]{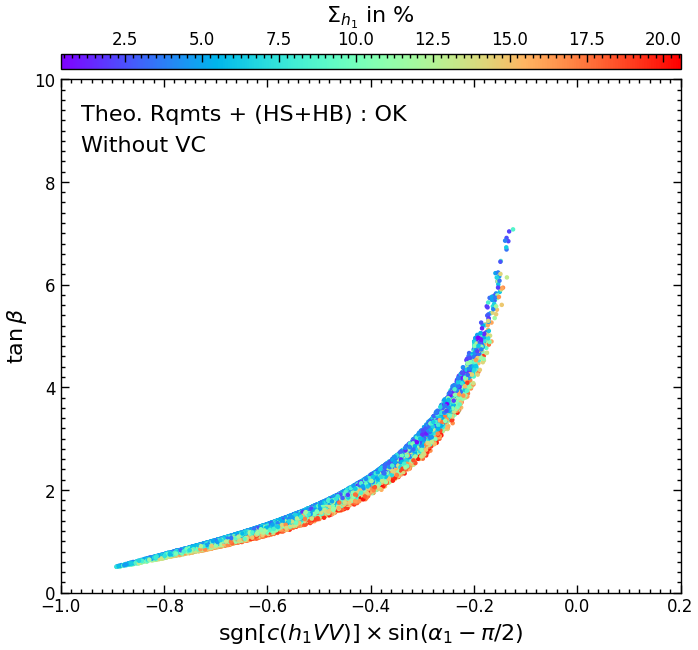} 
\includegraphics[width=0.3115\textwidth]{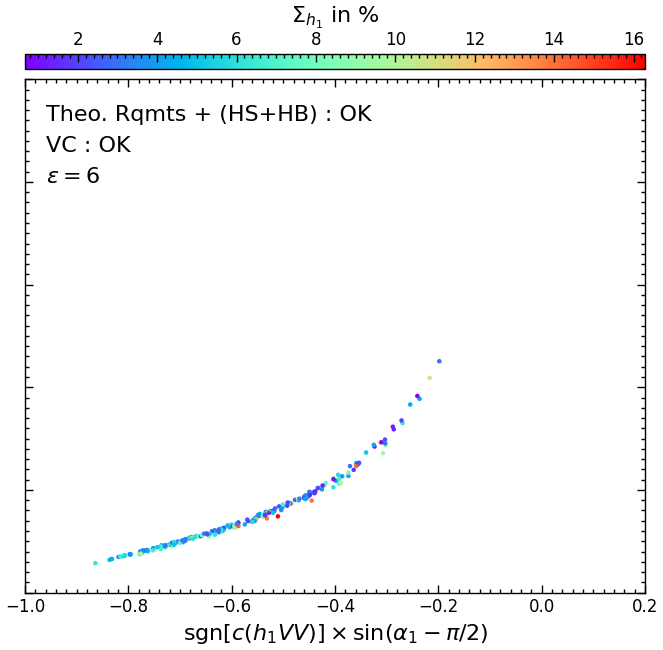} 
\includegraphics[width=0.3115\textwidth]{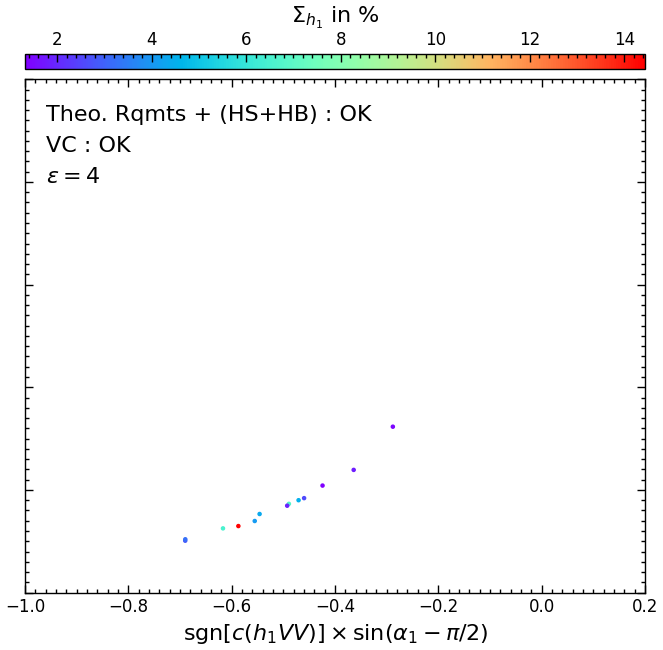}
\caption{The correct sign limit in the N2HDM as a function of the singlet admixture.}
\label{fig:condition-tb-correlation}
\end{figure}

To delve deeper into the results obtained, notably relating to mixing angles, we highlight in Fig.~\ref{fig:condition-tb-correlation} a scatter plot in the plane $\tan\beta$ and $\text{sgn}[c(h_1VV)]\times\sin(\alpha_1-\pi/2)$, quantifying the singlet contribution in the $h_1$ SM-like Higgs boson composition, which is defined by: $\Sigma_{h_1}=\big|\mathcal{R}_{13}\big|^2$. Importantly, upon incorporating the latest results from LHC Run II using HiggsTools, it becomes apparent that the parameter space of the N2HDM is constrained, and the wrong-sign regime (where the couplings of the $h_1=h_{125}$ to the fermions and massive gauge bosons are of opposite sign each other) seems to completely disappear. At this stage, significant singlet admixtures of up to $20\%$ are still compatible with the LHC Higgs data as can be seen from the left panel in Fig.~\ref{fig:condition-tb-correlation}.

Nevertheless, the situation about respect of naturalness considerations is more intriguing. In fact, including the VC will dramatically affect the so-called correct-sign regime that begins to shrink, and obviously, smaller values for the $\epsilon$-parameter result in a very limited area in the space parameter of N2HDM. The plots in Fig.~\ref{fig:condition-tb-correlation} show the corresponding generated samples. A prime example of this is reflected for $\epsilon=4$, where only $\tan\beta<3.2$ and $\sin\alpha_1$ delineated by the interval $[0.73;\,0.96]$ comply with all constraints. Furthermore, as can already be inferred from the middle and left plots, the singlet admixture is mostly restricted by the quasi-annulation of the VC, down from $20\%$ to about $14\%$ for $\epsilon=4$. It is beyond doubt, however, that such a proportion might be affected by any future measurements either at the HL-LHC or an electron-positron collider.

\subsection{Benchmark datasets}
\label{subsec:bps-n2hdm}
In this section, we investigate the convergence of the N2HDM towards the limit of a 2HDM decoupled from the singlet field. Such a situation may occur when $\alpha_1$ and $\alpha_{2,3}$ approach, $\alpha + \frac{\pi}{2}$ and $0$, respectively, where $\alpha$ represents the mixing angle used for diagonalizing the CP-even Higgs sector in the 2HDM. Consequently, the reduced couplings of the light Higgs boson, assumed to be the observed 125 GeV Higgs boson, to either $W^+W^-$ or $ZZ$ are perfectly aligned between the 2HDM and N2HDM. These couplings are as follows:
\begin{equation}
\label{eq:ChVV-2hdm-n2hdm}
g_{h_1VV}^{\text{N2HDM}} = c_{\alpha_2} c_{\beta-\alpha_1}   \xrightarrow[\text{convergence}]{}   g_{hVV}^{\text{2HDM}} = s_{\beta-\alpha}.
\end{equation}

\begin{figure}[!h]
\centering
\includegraphics[width=0.6\textwidth]{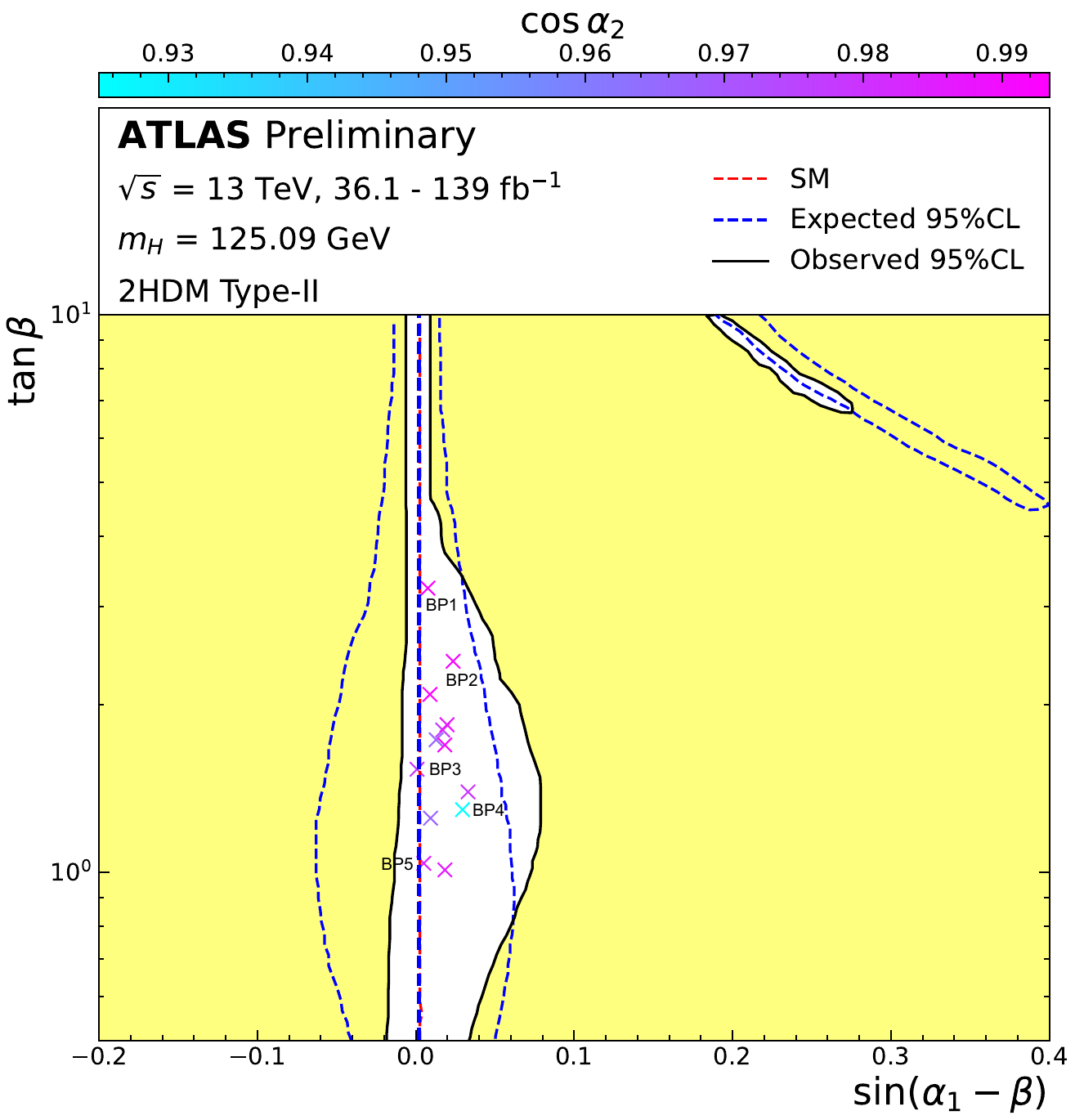} 
\caption{The type-II N2HDM allowed sample meeting Veltman conditions for $\epsilon=4$, marked by `×', is overlaid on the regions of the 2HDM Model II excluded by to the measured rates of Higgs boson production and
decays by  ATLAS \cite{ATLAS:2021vrm}. The color coding indicates for $\cos\alpha_2$ while the x-axis represents $\sin(\alpha_1-\beta)$ ($\cos(\beta-\alpha)$ in the 2HDM). The solid (short dashed) lines correspond to the observed (expected) limits at $2\sigma$. while the long-dashed vertical line indicates the SM prediction.}
\label{fig:sba1-tb-projection}
\end{figure}

\noindent
Hence, as in the 2HDM alignment limit: $c_{\beta-\alpha} =0 \iff s_{\beta-\alpha} =1$, one can proceed in exactly the same way within the N2HDM, and therefore:
\begin{equation}
\label{eq:ChVV-2hdm-n2hdm}
g_{h_1VV}^{\text{N2HDM}} =1   \iff  s_{\beta-\alpha_1}=0 \,\, \text{or} \,\,  s_{\alpha_2} =0.
\end{equation}
In light of these postulates, we depict in Fig.~\ref{fig:sba1-tb-projection} a projection of the surviving samples of the type-II N2HDM, while respecting the Veltman conditions for $\epsilon=4$, in the $\big[\sin(\beta-\alpha_1),\,\tan\beta\big]$ plane. Such a projection is overlaid on the expected exclusion limits at $2\sigma$ in the type-II 2HDM hypothesis, according to the previous convergence.
As can be seen, the model outcome falls squarely within the observed range at $2\sigma$, fully reflecting the consistency of N2HDM with the experimental measurements, together with the VC. Nonetheless, it's worth mentioning that for all the allowed samples, $\chi_V^{h_1}\times\chi_b^{h_1}$ is defined as positive. Therefore, only the correct-sign regime is included within the tested area, while the wrong-sign regime is typically excluded due to the alignment limits we impose.

Based on this result, we have selected five Benchmark Points (BPs), whose corresponding model inputs are given in Tab.~\ref{Th1-sm}. Additionally, we outline in Tab.~\ref{Th1-sm-2} the specifications of these benchmark points for the assumed 125-GeV Higgs boson, including the relative couplings to fermions and gauge bosons, i.e., $\chi_t^{h_1}$, $\chi_b^{h_1}$, $\chi_V^{h_1}$, the measured signal strengths as well as the $S$, $T$ and $U$ parameters. Thus, small $\tan\beta$ solutions, (below 3.5) exist, leading to a consistent agreement of $R_{\gamma \gamma}^{h_1}$ and $R_{Z\gamma}^{h_1}$ with existing data.

\begin{table*}[!h]
\scalebox{1.0}
{\begin{tabular}{|c|c|c|c|c|c|c|c|c|c|c|c|}
\hline
BPs &$t_\beta$ & $c_{\alpha_2}$ & $s_{\beta-\alpha_1}$ & $c_{\beta-\alpha_1}$ & $m_{h_1}$ & $m_{h_2}$ & $m_{h_3}$ & $m_A$ & $m_{H^{\pm}}$ & $\mu$ & $v_S$ \\
\hline
\hline
BP1 & 3.23 & 0.9927 & -0.0076  & 0.999971 & 125.09  & 676.714 & 779.429 & 700.329	& 708.470 & 363 & 379 \\ 
\hline 
BP2 & 2.39 & 0.9908 & -0.023606 & 0.999721 & 125.09 & 811.969 & 966.002 & 858.407 & 826.424 & 500.70 & 379.55 \\ 
\hline 
BP3 & 1.53 & 0.9778 & -0.000842 & 1.0 & 125.09 & 606.698 & 851.301 & 794.755 & 740.410 & 503.47 & 379.55 \\ 
\hline
BP4 & 1.29 & 0.9249 & -0.02957 & 0.999563 & 125.09 & 394.440 & 691.753 & 497.419 & 592.647 & 358.012 & 379.55 \\ 
\hline 
BP5 & 1.03 & 0.9835 & -0.005086 & 0.999987 & 125.09 & 601.137 & 696.009 & 661.480 & 643.149 & 459.600 & 379.55 \\ 
\hline
\end{tabular}}
\caption{Higgs bosons masses, $\mu$-parameter and singlet's {\it vev} $v_S$ (in GeV) are shown in the $h_1$-scenario for various values of angles $\alpha$'s and $\beta$.}
\label{Th1-sm}
\end{table*}

\begin{table*}[!h]
\scalebox{1.0}
{\begin{tabular}{|c|c|c|c|c|c|c|c|c|c|c|}
\hline 
 BPs &$ \chi_t^{h_1}$ & $\chi_b^{h_1}$ & $\chi_V^{h_1}$ & $R^{h_1}_{\gamma\gamma}$& $R^{h_1}_{Z\gamma} $  &  $\frac{\Gamma_{h_1}^{tot}}{\Gamma_{h}^{tot}(\text{SM})}$ &  $S$  &  $T$  &  $U$ \\ 
\hline
exp & $1.02^{+0.19}_{-0.15}$  & $0.91^{+0.17}_{-0.16}$  & $1.035^{+0.031}_{-0.031}$ & $1.04^{+0.1}_{-0.09}$ & $2.2^{+0.7}_{-0.7}$ & $ 0.98^{+0.31}_{-0.25}$ & $-0.02^{+0.1}_{-0.1}$ & $0.03^{+0.12}_{-0.12}$ & $0.01^{+0.11}_{-0.11}$\\
\hline
 BP1& 0.995 & 0.968 & 0.992& 1.032 & 1.027 & 0.955 &0.000502& 0.000621 & -0.000058 \\
\hline
 BP2 & 1.000 & 0.934 & 0.990& 1.076 & 1.081 & 0.913 &0.001837& 0.003784 & -0.000074 \\
\hline
 BP3 & 0.978 & 0.976 & 0.977 & 0.958 & 0.958 & 0.955 & 0.007983  & 0.024267 & -0.000128 \\
\hline 
 BP4 & 0.945 & 0.889 & 0.924 & 0.955 & 0.948 & 0.817 & 0.007795 & 0.026440 & -0.000193 \\
\hline
 BP5 & 0.988 & 0.978 & 0.983 & 0.980 & 0.982& 0.962 & 0.004677 & -0.003014 & -0.000128 \\
\hline
\end{tabular}} 
\caption{The $h_1$ SM-like relative couplings, decays rates and $S$, $T$ and $U$ parameters. The experimental data for $\chi_t^{h_1}$, $\chi_b^{h_1}$, $\chi_V^{h_1}$ \cite{CMS:2018uag}, $R^{h_1}_{\gamma\gamma}$ from \cite{ATLAS:2022tnm}, $R^{h_1}_{Z\gamma} $ from \cite{ATLAS:2023yqk}, for $\frac{\Gamma_{h_1}^{tot}}{\Gamma_{h_1}^{{tot}_{SM}}}$ from \cite{CMS:2018uag} and for oblique parameters from \cite{ParticleDataGroup:2022pth} are presented. }
\label{Th1-sm-2}
\end{table*}

Regarding charged and neutral Higgs boson masses, the VC solution exhibits other predictable features. It can be argued, for instance, that the two partners $h_2$ and $h_3$ are well controlled and their values lie respectively in $\sim 394-812$ GeV and $\sim 691-966$ GeV, while for the CP-odd Higgs boson, $A$, quasi-cancellation of quadratic divergences imperatively demands that $497 \lesssim m_A \lesssim 858$ GeV. The charged Higgs boson, meanwhile, exhibits a lower limit moderately higher than that imposed by the B-physics constraints, while its upper bounds have slightly decreased to 957 GeV.

Another interesting feature we can see in Fig.~\ref{fig:sba1-tb-projection} relates to the placement of our BPs in relation to the experimental best-fit value. Indeed, due to the limited sensitivity to the $\tan\beta$, the observed best-fit values for $c_{\beta-\alpha}$ is found to be $0.002$ in Type-II 2HDM, as reported by  ATLAS \cite{ATLAS:2021vrm}. Hence, considering the above convergence, it's obvious that BP3 and BP5 are close to this value and correspond to heavy $h_3$, $A$ and $H^{\pm}$ with a mass splitting ranging from $53$ GeV to $110$ GeV. Further to this, in view of the upcoming HL-LHC, all these benchmarks match well with the expected experimental results, particularly for the di-photon and Z+photon decays,  signal strengths, whose signal strengths read \cite{Cepeda:2019klc}:
\[ \mu_{\gamma\gamma}^{\text{HL-LHC}} = 1\pm0.04, \quad \mu_{Z\gamma}^{\text{HL-LHC}} = 1\pm0.23.  \]

Before ending, and given the significance of selected scenario, it is worth looking at the properties of the partners $h_2$ and $h_3$. Tables.\ref{Th2-partner} and \ref{Th3-partner} outline key properties of these heavy neutral Higgs bosons that are congruent with the cancellation of quadratic divergencies. Here, the couplings of the partners to quarks and gauge bosons are given by the SM prediction rescaled by a common factor:  $\chi_t^{h_a}$, $\chi_b^{h_a}$ and $\chi_V^{h_a}$ (a=2,3), evaluated with respect to couplings the would-be SM Higgs boson with the same mass as $h_a$. 

Accordingly, for $h_2$, Tab.~\ref{Th2-partner} makes it clear that the corresponding coupling to down-quarks is negative, which is easily explained by looking at the $h_2 d \bar{d}$ coupling former (see Tab.~\ref{Type2-Hiff-HiVV}), especially that for all the survived points the mixing angle $\alpha_2$ (resp. $\alpha_3$) values range from -0.38 (resp. -0.9) to -0.12 (resp. 0). Meanwhile, its coupling $\chi_V^{h_2}$ has diminished for all BPs compared to the SM value. Furthermore, significant increase of $R^{h_2}_{\gamma\gamma}$ was observed, ranging from 0.089 to 60 times larger the SM one, contrary to what $R^{h_2}_{Z\gamma}$ which looks suppressed overall. The last column in Tab.~\ref{Th2-partner} indicates that the total width $\Gamma_{h_2}^{tot}$ remains far below its SM value; and as the decay into $ZZ$ is a nearly zero the di-photon appears to be the most promising channel in searching for such scalar. Moreover, as $m_{h_2} < m_{A} + m_{Z}$, $m_{H^\pm} + m_{W^\pm}$, $2 m_{A}$ for all survived points in Fig.~\ref{fig:sba1-tb-projection}, decays such as $h_2 \to A Z$, $H^\pm W^\mp$, and $A A$ are suppressed. Only the $h_2 \to h_1 h_1$ decay channel is found to be competitive with $VV$, especially $W^+W^-$, and reaches $43\%$ for $m_{h_2} \sim 394$ GeV before the threshold where the decay into a pair of top quarks is kinematically open and dominates at high values of $m_{h_2}$.

\begin{table}[!h]
\scalebox{1.0}
{\begin{tabular}{|c|c|c|c|c|c|c|c|}
\hline 
 BPs & $m_{h_2}$ & $\chi_t^{h_2}$ & $\chi_b^{h_2}$ & $\chi_V^{h_2}$ & $R^{h_2}_{\gamma\gamma}$ & $R^{h_2}_{Z\gamma}$ & $\frac{\Gamma_{h_2}^{tot}}{\Gamma_{h}^{tot}(\text{SM})}$ \\
\hline
\hline
BP1 & 676.714 & 0.257 & -3.151 & -3.99$\times10^{-2}$ & 12.282 & 0.093 & 0.013 \\
\hline
BP2 & 811.969 & 0.306 & -2.260 & -7.59$\times10^{-2}$ & 2.623 & 0.204 & 0.021 \\
\hline
BP3 & 606.698 & 0.197 & -1.040 & -0.173 & 4.996 & 0.065 & 0.044 \\
\hline
BP4 & 394.440 & 0.196 & -1.148 & -0.305 & 0.089 & 0.032 & 0.155 \\
\hline
BP5 & 601.137 & 0.523 & -0.844 & -0.135 & 60.370 & 0.407 & 0.069 \\ 
\hline
\end{tabular}} 
\caption{The mass, relative couplings and decays rates for the partner $h_2$ Higgs boson.}
\begin{center}
\label{Th2-partner}
\end{center} 
\end{table}
\begin{table*}[!h]
\scalebox{1.0}
{\begin{tabular}{|c|c|c|c|c|c|c|c|}
\hline 
 BPs & $m_{h_3}$ & $\chi_t^{h_3}$ & $\chi_b^{h_3}$ & $\chi_V^{h_3}$ & $R^{h_3}_{\gamma\gamma}$ & $R^{h_3}_{Z\gamma}$ & $\frac{\Gamma_{h_3}^{tot}}{\Gamma_{h}^{tot}(\text{SM})}$ \\
\hline
\hline
BP1 & 779.429 & 0.197 & -0.760 & 0.114 & 0.907 & 0.022 & 0.010 \\
\hline
BP2 & 966.002 & 0.283 & -0.855 & 0.114 & 0.090 & 0.010 & 0.105 \\
\hline
BP3 & 851.301 & 0.656 & -1.143 & 0.118 & 1.607 & 0.082 & 0.349 \\
\hline
BP4 & 691.753 & 0.813 & -0.755 & 0.228 & 4.538 & 0.025 & 1.962 \\
\hline
BP5 & 696.009 & 0.822 & -0.638 & 0.119 & 81.666 & 0.506 & 0.106 \\ 
\hline
\end{tabular}} 
\caption{The mass, relative couplings and decays rates for the partner $h_3$ Higgs boson.}
\begin{center}
\label{Th3-partner}
\end{center} 
\end{table*}

And lastly, the heavier partner ($h_3$) exhibits roughly the same features, as can be seen from Tab.~\ref{Th3-partner}. The signal strength $R^{h_3}_{\gamma\gamma}$ shows a significant enhancement, amounting to 81 times its SM value, in contrast to $R^{h_3}_{Z\gamma}$ which remains suppressed overall similar to $h_2$. Meanwhile, the reduced total width $\Gamma_{h_3}^{tot}$ may have further improvement-the corresponding ratio ranges from 0.01 to 1.96. On the other hand, despite its relative couplings to down-quarks being modest (-1.14 up to -0.63), the $h_3$ decay into $b\bar{b}$ is seen to be suppressed, with the $t\bar{t}$-channel being dominant. Thus, from a value of $90\%$ ($m_{h_3} \sim 600$ GeV), the branching ratio $Br(h_3 \to t\bar{t})$ declines towards $10\%$ for higher values of $m_{h_3}$. This is because the $h_3 \to h_1 h_2$ and $h_3 \to A Z$ channels open up, and the involved couplings, i.e, $c_{h_1 h_2 h_3}$ and $c_{Z A h_3}$ are non zero and even enhanced as $\sin(\alpha_1-\beta)$ gets closer to zero. Note that the Higgs-to-Higgs decay $h_3 \to h_1 h_2$ dominates for higher mass, $m_{h_3} \gtrsim 851$ GeV, and may be useful as soon as $m_{h_2} > 250$ GeV, where the $h_2 \to h_1 h_1$ channel becomes dominant, quickly reaching a rate of $\sim33\%$. As a result, an intermediate $h_1 h_1 h_1$ state can in general be sizable and still pose a challenge for the HL-LHC, given its dominant direct decay modes to SM particles.

\section{Conclusion}
\label{sec:conlusion}
We have explored in this study to what extent the cancellation of quadratic divergences may theoretically restrict the parameter space of the N2HDM. In line with positivity and perturbativity requirements, we have considered the $h_1$-scenario where the lightest CP-even Higgs boson is assumed to be the observed 125 GeV. Furthermore, we have calculated the quadratic divergent correction for the Higgs boson self-energy, e.g., $h_1$ and its partners $h_2$ and $h_3$, and demanded that those quantities be made zero or at least controllably small.

Then we have selected benchmark points of the type II N2HDM parameter space in the $\sin(\alpha_1-\beta)$ and the $\tan\beta$ plane, where our results showed that canceling out the one-loop quadratic divergences in the Higgs mass leaves a viable and significant parameter region, conducive to probing NP, while retaining consistency with the HL-LHC predictions, as well as the oblique parameters $S$, $T$ and $U$, at the $2\sigma$ level.

Signal strength measurements of the SM-like Higgs boson $h_1$ and its partners $h_2$ and $h_3$ indicate interesting prospects within the type II N2HDM framework. While certain decay channels may be inaccessible or suppressed, our findings underscore a potential challenges for exploring physics BSM.

\section*{Acknowledgements}
The authors would like to thank Michel Capdequi Peyran\`ere for useful discussions. This work is supported by the Moroccan Ministry of Higher Education and Scientific Research MESRSFC and CNRST: Projet PPR/2015/6. SS is fully supported through the NExT Institute.

\appendix

\section{Theoretical Constraints}
\label{appdice:theoconstraints}
In this section we recall two important theoretical constraints. Firstly, we start with the unitarity, which derives new restrictions on the Higgs and potential parameters in the N2HDM, by ensuring that tree-level perturbative unitarity is preserved. For more details, we refer the reader to Ref. \cite{Arhrib:2018qmw}, which provides the explicit calculations for this purpose.  The corresponding eigenvalues found at tree level are given by:
\begin{eqnarray}
|\lambda_3+\lambda_4|\ ,\ |\lambda_3\pm\lambda_5|\ , \ |\lambda_3+2\lambda_4 \pm 3\lambda_5| < 8\pi \nonumber \\
\big|\frac{\lambda_7}{2}\big|\ , \ \big|\frac{\lambda_8}{2}\big|\ , \ |\frac{\lambda_6}{4}|< 8\pi \nonumber \\
\big|\frac{1}{2}(\lambda_1+\lambda_2\pm\sqrt{(\lambda_1-\lambda_2)^2+4\lambda_4^2})\big|< 8\pi \nonumber \\
\big|\frac{1}{2}(\lambda_1+\lambda_2\pm\sqrt{(\lambda_1-\lambda_2)^2+4\lambda_5^2})\big|< 8\pi \nonumber
\end{eqnarray}
are derived with others ones.

Secondly, and for the scalar potential to be bounded from below in all directions as the fields approach infinity, the boundedness from below (BFB) conditions must be satisfied. In the N2HDM case, where there is no connection between the three fields, the conditions $\lambda_1>0$, $\lambda_2>0$, and $\lambda_6>0$ are sufficient to ensure this requirement. Additionally, while picking up all ﬁeld space directions, the remaining BFB constraints depends on the discriminant 
\beq
D = \left\{ \begin{array}{lll} \lambda_4 - \lambda_5 & \mbox{for} &
    \lambda_4 > \lambda_5 \\ 0 & \mbox{for} & \lambda_4 \le \lambda_5 \end{array}\right.,
\eeq
and reads as \cite{Muhlleitner:2016mzt,Arhrib:2018qmw}
\beq
\Omega_1 \cup \Omega_2
\eeq
where
\beq
&& \Omega_1 = \Bigg\{ \sqrt{\lambda_1 \lambda_6} +
\lambda_7 > 0,\, \sqrt{\lambda_2 \lambda_6} + \lambda_8 > 0,\, \sqrt{\lambda_1 \lambda_2} + \lambda_3 + D > 0,\, \lambda_7 +
\lambda_8\sqrt{\lambda_1/\lambda_2} \ge 0 \Bigg\}, \label{eq:omega1} \nonumber\\
&& \Omega_2 = \Bigg\{\sqrt{\lambda_2 \lambda_6} \ge
\lambda_8 > -\sqrt{\lambda_2 \lambda_6},\, \sqrt{\lambda_1 \lambda_6} > -
\lambda_7 \ge \lambda_8\sqrt{\lambda_1/\lambda_2} ,\nonumber\\
&& \hspace{6.8cm} \sqrt{(\lambda_7^2 - \lambda_1 \lambda_6)(\lambda_8^2 -\lambda_2
\lambda_6)} > \lambda_7 \lambda_8 - (D+\lambda_3) \lambda_6 \Bigg\}
\label{eq:omega2}
\nonumber
\eeq

\section{Feynman rules}
\label{appdice:feynmanRules}
In this appendix, we provide the necessary couplings for computing the tadpoles formulas of each CP-even Higgs boson, denoted as $T_{h_1}$, $T_{h_2}$ and $T_{h_3}$ considering only the three-leg couplings involved at the one-loop contributions. For such purpose, we have adopted the well-known linear $R_{\xi}$ gauge, with gauge-fixing Lagrangians given by $\frac{-1}{2\xi_Z} (\partial_{\mu} Z^{\mu} -\xi_Z m_Z G_0)^2$ for the neutral sector and $\frac{-1}{2\xi_W}(\partial_{\mu} W_{\pm}^{\mu} -\xi_W m_W G_{\pm})^2$ for the charged sector, and investigated all vertices such as $h_{i} F_{j} \bar {F_j}$ (i=1,2,3), where $F_j$ represents any quantum field in our model, including scalar, vector bosons, fermions, Goldstone fields $G_k$, and Faddeev-Popov ghost fields $\eta_k$. Tab.\ref{tab:tadpoleh1} brings together all the couplings $c_i^{h_1}$, symmetry factor $s_i^{h_1}$ and the propagator loops $t_i^{h_1}$, used for the $T_{h_1}$ calculation, while the remaining Tab.\ref{tab:tadpoleh2} and Tab.\ref{tab:tadpoleh3} provide the same inputs for the $T_{h_{2,3}}$ ones. 
\setlength{\arrayrulewidth}{0.4mm}
\afterpage{
\begin{sidewaystable}[!h]
\centering 
\small
\begin{tabular}{|l|l|c|c|}
\hline 
\multicolumn{4}{|c|}{Tadpole calculations for $h_1$ : $T_{h_1}$} \\
\hline 
\multicolumn{1}{|c|}{\text{vertex}} & \multicolumn{1}{c|}{\text{coupling} $c_i^{h_1}$} & \multicolumn{1}{c|}{$s_i^{h_1}$} & \multicolumn{1}{c|}{$t_i^{h_1}$} \\
\hline 
\hline
$h_1$-$h_1$-$h_1$ &
  \begin{tabular}[t]{@{}l@{}}
	$-\frac{3}{4}i \big[ 2c_\beta \mathcal{R}_{11} \left(2 \mathcal{R}_{11}^2\lambda_1 + 2 \mathcal{R}_{12}^2\lambda_{345} + \mathcal{R}_{13}^2\lambda_7 \right)v
	+ 2\mathcal{R}_{12} \left(2\mathcal{R}_{12}^2\lambda_2 + 2\mathcal{R}_{11}^2\lambda_{345} + \mathcal{R}_{13}^2\lambda_8\right) s_\beta v$ \\
	$+ \mathcal{R}_{13}\left(\mathcal{R}_{13}^2 \lambda_6 + 2\mathcal{R}_{11}^2 \lambda_7 + 2\mathcal{R}_{12}^2\lambda_8\right)v_s \big]$
  \end{tabular} & 1/2 & $i A_0(m_{h_1}^2)$ \\
\hline
$h_1$-$h_2$-$h_2$ & 
  \begin{tabular}[t]{@{}l@{}}
    $-\frac{1}{4}i \big[ 2 \left(2\mathcal{R}_{21} (2\mathcal{R}_{12}\mathcal{R}_{22}\lambda_{345} + \mathcal{R}_{13}\mathcal{R}_{23}\lambda_{7}) + \mathcal{R}_{11}(6\mathcal{R}_{21}^2\lambda_1 + 2\mathcal{R}_{22}^2\lambda_{345} + \mathcal{R}_{23}^2\lambda_7)\right)c_\beta v $ \\
    $+4\left( \mathcal{R}_{21}(2\mathcal{R}_{11}\mathcal{R}_{22} + \mathcal{R}_{12}\mathcal{R}_{21})\lambda_{345} + \mathcal{R}_{23}(\mathcal{R}_{13}\mathcal{R}_{22} + \frac{1}{2} \mathcal{R}_{12}\mathcal{R}_{23})\lambda_8 + 3\mathcal{R}_{12}\mathcal{R}_{22}^2\lambda_2\right)s_\beta v $ \\
    $+ \left(3\mathcal{R}_{13}\mathcal{R}_{23}^2\lambda_6 + 2\mathcal{R}_{21}(\mathcal{R}_{13}\mathcal{R}_{21} + 2\mathcal{R}_{11}\mathcal{R}_{23})\lambda_7 + 2\mathcal{R}_{22}(\mathcal{R}_{13}\mathcal{R}_{22} + 2\mathcal{R}_{12}\mathcal{R}_{23})\lambda_8\right)v_s \big]$ 
  \end{tabular} & 1/2 & $i A_0(m_{h_2}^2)$ \\    
\hline
$h_1$-$h_3$-$h_3$ & 
  \begin{tabular}[t]{@{}l@{}}
   $-\frac{1}{4}i \big[ 2 \left(2\mathcal{R}_{31} (2\mathcal{R}_{12}\mathcal{R}_{32}\lambda_{345} + \mathcal{R}_{13}\mathcal{R}_{33}\lambda_{7}) + \mathcal{R}_{11}(6\mathcal{R}_{31}^2\lambda_1 + 2\mathcal{R}_{32}^2\lambda_{345} + \mathcal{R}_{33}^2\lambda_7)\right)c_\beta v $ \\
   $+4\left( \mathcal{R}_{31}(2\mathcal{R}_{11}\mathcal{R}_{32} + \mathcal{R}_{12}\mathcal{R}_{31})\lambda_{345} + \mathcal{R}_{33}(\mathcal{R}_{13}\mathcal{R}_{32} + \frac{1}{2} \mathcal{R}_{12}\mathcal{R}_{33})\lambda_8 + 3\mathcal{R}_{12}\mathcal{R}_{32}^2\lambda_2\right)s_\beta v $ \\
   $+ \left(3\mathcal{R}_{13}\mathcal{R}_{33}^2\lambda_6 + 2\mathcal{R}_{31}(\mathcal{R}_{13}\mathcal{R}_{31} + 2\mathcal{R}_{11}\mathcal{R}_{33})\lambda_7 + 2\mathcal{R}_{32}(\mathcal{R}_{13}\mathcal{R}_{32} + 2\mathcal{R}_{12}\mathcal{R}_{33})\lambda_8\right)v_s \big]$	  	
  \end{tabular} & 1/2 & $i A_0(m_{h_3}^2)$ \\
\hline  
$h_1$-$A$-$A$ & 
  \begin{tabular}[t]{@{}l@{}}
   $-\frac{1}{2}i \big[ 2c^3_\beta \mathcal{R}_{11}\left(\lambda_3+\lambda_4-\lambda_5\right)v + 2c_\beta \mathcal{R}_{11}\left(\lambda_1-2\lambda_5\right)s^2_\beta v 
   + s^2_\beta \left(2\mathcal{R}_{12} (\lambda_3+\lambda_4-\lambda_5)s_\beta v + \mathcal{R}_{13}\lambda_7 v_s  \right) $ \\
   $+ c^2_\beta \left( 2\mathcal{R}_{12}(\lambda_2-2\lambda_5)s_\beta v + \mathcal{R}_{13} \lambda_8 v_s\right) \big]$           
  \end{tabular} & 1/2 & $i A_0(m_{A}^2)$ \\
\hline              
$h_1$-$G$-$G$ & 
  \begin{tabular}[t]{@{}l@{}}     
   $-\frac{1}{2}i \big[ 2c^3_\beta \mathcal{R}_{11} \lambda_1 v + 2c_\beta \mathcal{R}_{11} \lambda_{345} s^2_\beta v 
   + s^2_\beta \left(2\mathcal{R}_{12} \lambda_2 s_\beta v + \mathcal{R}_{13}\lambda_8 v_s  \right)  
   + c^2_\beta \left(2\mathcal{R}_{12} \lambda_{345} s_\beta v + \mathcal{R}_{13} \lambda_7 v_s\right) \big]$	   	                   
  \end{tabular} & 1/2 & $i A_0(\xi_Z m_Z^2)$ \\
\hline  
$h_1$-$H^+$-$H^-$ &
  \begin{tabular}[t]{@{}l@{}}
   $-\frac{1}{2}i \big[ 2c^3_\beta \mathcal{R}_{11} \lambda_3 v + 2c_\beta \mathcal{R}_{11} \left(\lambda_1-\lambda_4-\lambda_5\right) s^2_\beta v 
   + s^2_\beta \left(2\mathcal{R}_{12} \lambda_3 s_\beta v + \mathcal{R}_{13}\lambda_7 v_s  \right) $ \\
   $+ c^2_\beta \left(2\mathcal{R}_{12} \left(\lambda_2-\lambda_4-\lambda_5\right) s_\beta v + \mathcal{R}_{13} \lambda_8 v_s\right) \big]$  	
  \end{tabular} & 1/2 & $2 \, \big[ i A_0(m_{H^\pm}^2)\big]$ \\
\hline
$h_1$-$G^+$-$G^-$ &
  \begin{tabular}[t]{@{}l@{}}
   $-\frac{1}{2}i \big[ 2c^3_\beta \mathcal{R}_{11} \lambda_1 v + 2c_\beta \mathcal{R}_{11} \lambda_{345} s^2_\beta v 
   + s^2_\beta \left(2\mathcal{R}_{12} \lambda_2 s_\beta v + \mathcal{R}_{13}\lambda_8 v_s  \right) 
   + c^2_\beta \left(2\mathcal{R}_{12} \lambda_{345} s_\beta v + \mathcal{R}_{13} \lambda_7 v_s\right) \big]$  	
  \end{tabular}  & 1/2 & $2 \, \big[i A_0(\xi_W m_W^2)\big]$ \\
\hline
$h_1$-$z$-$z$ & 
  \begin{tabular}[t]{@{}l@{}}
   $ ig m_W \left(c_\beta \mathcal{R}_{11} + s_\beta \mathcal{R}_{12} \right) / c^2_w $
  \end{tabular}  & 1/2 & $-i\big[(n-1) A_0(m_Z^2) + \xi_Z A_0(\xi_Z m_Z^2)\big]$ \\
\hline
$h_1$-$w^+$-$w^-$ &
  \begin{tabular}[t]{@{}l@{}}
   $ ig m_W \left(c_\beta \mathcal{R}_{11} + s_\beta \mathcal{R}_{12} \right)  $
  \end{tabular}  & 1/2 & $-2i \, \big[(n-1) A_0(m_W^2) + \xi_W A_0(\xi_W m_W^2)\big]$ \\
\hline
$h_1$-$u$-$\bar{u}$ & 
  \begin{tabular}[t]{@{}l@{}}
   $-\frac{i}{2} g \left(\mathcal{R}_{12}/ s_\beta \right) m_U / m_W   $  	
  \end{tabular}  & 1 & $-i \, m_{U} \, A_0(m_{U}^2) \,\text{Tr}(I_n)$  \\
\hline
$h_1$-$d$-$\bar{d}$ &
  \begin{tabular}[t]{@{}l@{}}
	$-\frac{i}{2} g \left(\mathcal{R}_{11}/ c_\beta \right) m_D / m_W   $
  \end{tabular}  & 1 & $-i \, m_{D} \, A_0(m_{D}^2) \,\text{Tr}(I_n)$   \\
\hline
$h_1$-$\eta_Z$-$\bar{\eta_Z}$ & 
  \begin{tabular}[t]{@{}l@{}}
    $ -\frac{i}{2} g m_W \left(c_\beta \mathcal{R}_{11} + s_\beta \mathcal{R}_{12} \right) \xi_Z  / c^2_w  $ 
  \end{tabular}  & 1 & $i \, A_0(\xi_Z m_Z^2)$   \\
\hline
$h_1$-$\eta_\pm$-$\bar{\eta_\pm}$ &
  \begin{tabular}[t]{@{}l@{}}
	$ -\frac{i}{2} g m_W \left(c_\beta \mathcal{R}_{11} + s_\beta \mathcal{R}_{12} \right) \xi_W   $          
  \end{tabular}  & 1 & $2i \, A_0(\xi_W m_W^2)$  \\
\hline
\end{tabular}
\caption{The couplings $c_i^{h_1}$, symmetry factor $s_i^{h_1}$ and the propagator loops $t_i^{h_1}$ used for $T_{h_1}$ calculation.}
\label{tab:tadpoleh1}
\end{sidewaystable} 
\clearpage
}

\setlength{\arrayrulewidth}{0.4mm}
\afterpage{
	\begin{sidewaystable}[!h]
		\centering 
		\small
		\begin{tabular}{|l|l|c|c|}
			\hline 
			\multicolumn{4}{|c|}{Tadpole calculations for $h_2$ : $T_{h_2}$} \\
			\hline 
            \multicolumn{1}{|c|}{\text{vertex}} & \multicolumn{1}{c|}{\text{coupling} $c_i^{h_2}$} & \multicolumn{1}{c|}{$s_i^{h_2}$} & \multicolumn{1}{c|}{$t_i^{h_2}$} \\
			\hline 
			\hline
			$h_2$-$h_2$-$h_2$ &
			\begin{tabular}[t]{@{}l@{}}
				$-\frac{3}{4}i \big[ 2c_\beta \mathcal{R}_{21} \left(2 \mathcal{R}_{21}^2\lambda_1 + 2 \mathcal{R}_{22}^2\lambda_{345} + \mathcal{R}_{23}^2\lambda_7 \right)v
				+ 2\mathcal{R}_{22} \left(2\mathcal{R}_{22}^2\lambda_2 + 2\mathcal{R}_{21}^2\lambda_{345} + \mathcal{R}_{23}^2\lambda_8\right) s_\beta v$ \\
				$+ \mathcal{R}_{23}\left(\mathcal{R}_{23}^2 \lambda_6 + 2\mathcal{R}_{21}^2 \lambda_7 + 2\mathcal{R}_{22}^2\lambda_8\right)v_s \big]$
			\end{tabular} & 1/2 & $i A_0(m_{h_2}^2)$ \\
			\hline
			$h_2$-$h_1$-$h_1$ & 
			\begin{tabular}[t]{@{}l@{}}
				$-\frac{1}{4}i \big[ 2 \left(2\mathcal{R}_{11} (2\mathcal{R}_{12}\mathcal{R}_{22}\lambda_{345} + \mathcal{R}_{13}\mathcal{R}_{23}\lambda_{7}) + \mathcal{R}_{21}(6\mathcal{R}_{11}^2\lambda_1 + 2\mathcal{R}_{12}^2\lambda_{345} + \mathcal{R}_{13}^2\lambda_7)\right)c_\beta v $ \\
				$+4\left( \mathcal{R}_{11}(\mathcal{R}_{11}\mathcal{R}_{22} + 2\mathcal{R}_{12}\mathcal{R}_{21})\lambda_{345} + \mathcal{R}_{13}(\frac{1}{2} \mathcal{R}_{13}\mathcal{R}_{22} + \mathcal{R}_{12}\mathcal{R}_{23})\lambda_8 + 3\mathcal{R}_{12}^2\mathcal{R}_{22}\lambda_2\right)s_\beta v $ \\
				$+ \left(3\mathcal{R}_{13}^2\mathcal{R}_{23}\lambda_6 + 2\mathcal{R}_{11}(2\mathcal{R}_{13}\mathcal{R}_{21} + \mathcal{R}_{11}\mathcal{R}_{23})\lambda_7 + 2\mathcal{R}_{12}(\mathcal{R}_{12}\mathcal{R}_{23} + 2\mathcal{R}_{13}\mathcal{R}_{22})\lambda_8\right)v_s \big]$ 
			\end{tabular} & 1/2 & $i A_0(m_{h_1}^2)$ \\    
			\hline
			$h_2$-$h_3$-$h_3$ & 
			\begin{tabular}[t]{@{}l@{}}
				$-\frac{1}{4}i \big[ 2 \left(2\mathcal{R}_{31} (2\mathcal{R}_{22}\mathcal{R}_{32}\lambda_{345} + \mathcal{R}_{23}\mathcal{R}_{33}\lambda_{7}) + \mathcal{R}_{21}(6\mathcal{R}_{31}^2\lambda_1 + 2\mathcal{R}_{32}^2\lambda_{345} + \mathcal{R}_{33}^2\lambda_7)\right)c_\beta v $ \\
				$+4\left( \mathcal{R}_{31}(2\mathcal{R}_{21}\mathcal{R}_{32} + \mathcal{R}_{22}\mathcal{R}_{31})\lambda_{345} + \mathcal{R}_{33}(\mathcal{R}_{23}\mathcal{R}_{32} + \frac{1}{2} \mathcal{R}_{22}\mathcal{R}_{33})\lambda_8 + 3\mathcal{R}_{22}\mathcal{R}_{32}^2\lambda_2\right)s_\beta v $ \\
				$+ \left(3\mathcal{R}_{23}\mathcal{R}_{33}^2\lambda_6 + 2\mathcal{R}_{31}(\mathcal{R}_{23}\mathcal{R}_{31} + 2\mathcal{R}_{21}\mathcal{R}_{33})\lambda_7 + 2\mathcal{R}_{32}(\mathcal{R}_{23}\mathcal{R}_{32} + 2\mathcal{R}_{32}\mathcal{R}_{33})\lambda_8\right)v_s \big]$	  	
			\end{tabular} & 1/2 & $i A_0(m_{h_3}^2)$ \\
			\hline  
			$h_2$-$A$-$A$ & 
			\begin{tabular}[t]{@{}l@{}}
				$-\frac{1}{2}i \big[ 2c^3_\beta \mathcal{R}_{21}\left(\lambda_3+\lambda_4-\lambda_5\right)v + 2c_\beta \mathcal{R}_{21}\left(\lambda_1-2\lambda_5\right)s^2_\beta v 
				+ s^2_\beta \left(2\mathcal{R}_{22} (\lambda_3+\lambda_4-\lambda_5)s_\beta v + \mathcal{R}_{23}\lambda_7 v_s  \right) $ \\
				$+ c^2_\beta \left( 2\mathcal{R}_{22}(\lambda_2-2\lambda_5)s_\beta v + \mathcal{R}_{23} \lambda_8 v_s\right) \big]$           
			\end{tabular} & 1/2 & $i A_0(m_{A}^2)$ \\
			\hline              
			$h_2$-$G$-$G$ & 
			\begin{tabular}[t]{@{}l@{}}     
				$-\frac{1}{2}i \big[ 2c^3_\beta \mathcal{R}_{21} \lambda_1 v + 2c_\beta \mathcal{R}_{21} \lambda_{345} s^2_\beta v 
				+ s^2_\beta \left(2\mathcal{R}_{22} \lambda_2 s_\beta v + \mathcal{R}_{23}\lambda_8 v_s  \right)  
				+ c^2_\beta \left(2\mathcal{R}_{22} \lambda_{345} s_\beta v + \mathcal{R}_{23} \lambda_7 v_s\right) \big]$	   	                   
			\end{tabular} & 1/2 & $i A_0(\xi_Z m_Z^2)$ \\
			\hline  
			$h_2$-$H^+$-$H^-$ &
			\begin{tabular}[t]{@{}l@{}}
				$-\frac{1}{2}i \big[ 2c^3_\beta \mathcal{R}_{21} \lambda_3 v + 2c_\beta \mathcal{R}_{21} \left(\lambda_1-\lambda_4-\lambda_5\right) s^2_\beta v 
				+ s^2_\beta \left(2\mathcal{R}_{22} \lambda_3 s_\beta v + \mathcal{R}_{23}\lambda_7 v_s  \right) $ \\
				$+ c^2_\beta \left(2\mathcal{R}_{22} \left(\lambda_2-\lambda_4-\lambda_5\right) s_\beta v + \mathcal{R}_{23} \lambda_8 v_s\right) \big]$  	
			\end{tabular} & 1/2 & $2 \, \big[ i A_0(m_{H^\pm}^2)\big]$ \\
			\hline
			$h_2$-$G^+$-$G^-$ &
			\begin{tabular}[t]{@{}l@{}}
				$-\frac{1}{2}i \big[ 2c^3_\beta \mathcal{R}_{21} \lambda_1 v + 2c_\beta \mathcal{R}_{21} \lambda_{345} s^2_\beta v 
				+ s^2_\beta \left(2\mathcal{R}_{22} \lambda_2 s_\beta v + \mathcal{R}_{23}\lambda_8 v_s  \right) 
				+ c^2_\beta \left(2\mathcal{R}_{22} \lambda_{345} s_\beta v + \mathcal{R}_{23} \lambda_7 v_s\right) \big]$  	
			\end{tabular}  & 1/2 & $2 \, \big[i A_0(\xi_W m_W^2)\big]$ \\
			\hline
			$h_2$-$z$-$z$ & 
			\begin{tabular}[t]{@{}l@{}}
				$ ig m_W \left(c_\beta \mathcal{R}_{21} + s_\beta \mathcal{R}_{22} \right) / c^2_w $
			\end{tabular}  & 1/2 & $-i\big[(n-1) A_0(m_Z^2) + \xi_Z A_0(\xi_Z m_Z^2)\big]$ \\
			\hline
			$h_2$-$w^+$-$w^-$ &
			\begin{tabular}[t]{@{}l@{}}
				$ ig m_W \left(c_\beta \mathcal{R}_{21} + s_\beta \mathcal{R}_{22} \right)  $
			\end{tabular}  & 1/2 & $-2i \, \big[(n-1) A_0(m_W^2) + \xi_W A_0(\xi_W m_W^2)\big]$ \\
			\hline
			$h_2$-$u$-$\bar{u}$ & 
			\begin{tabular}[t]{@{}l@{}}
				$-\frac{i}{2} g \left(\mathcal{R}_{22}/ s_\beta \right) m_U / m_W   $  	
			\end{tabular}  & 1 & $-i \, m_{U} \, A_0(m_{U}^2) \,\text{Tr}(I_n)$  \\
			\hline
			$h_2$-$d$-$\bar{d}$ &
			\begin{tabular}[t]{@{}l@{}}
				$-\frac{i}{2} g \left(\mathcal{R}_{21}/ c_\beta \right) m_D / m_W   $
			\end{tabular}  & 1 & $-i \, m_{D} \, A_0(m_{D}^2) \,\text{Tr}(I_n)$   \\
			\hline
			$h_2$-$\eta_Z$-$\bar{\eta_Z}$ & 
			\begin{tabular}[t]{@{}l@{}}
				$ -\frac{i}{2} g m_W \left(c_\beta \mathcal{R}_{21} + s_\beta \mathcal{R}_{22} \right) \xi_Z  / c^2_w  $ 
			\end{tabular}  & 1 & $i \, A_0(\xi_Z m_Z^2)$   \\
			\hline
			$h_2$-$\eta_\pm$-$\bar{\eta_\pm}$ &
			\begin{tabular}[t]{@{}l@{}}
				$ -\frac{i}{2} g m_W \left(c_\beta \mathcal{R}_{21} + s_\beta \mathcal{R}_{22} \right) \xi_W   $          
			\end{tabular}  & 1 & $2i \, A_0(\xi_W m_W^2)$  \\
			\hline
		\end{tabular}
		\caption{The couplings $c_i^{h_2}$, symmetry factor $s_i^{h_2}$ and the propagator loops $t_i^{h_2}$ used for $T_{h_2}$ calculation.}
		\label{tab:tadpoleh2}
	\end{sidewaystable} 
	\clearpage
}

\setlength{\arrayrulewidth}{0.4mm}
\afterpage{
	\begin{sidewaystable}[!h]
		\centering 
		\small
		\begin{tabular}{|l|l|c|c|}
			\hline 
			\multicolumn{4}{|c|}{Tadpole calculations for $h_3$ : $T_{h_3}$} \\
			\hline 
			\multicolumn{1}{|c|}{\text{vertex}} & \multicolumn{1}{c|}{\text{coupling} $c_i^{h_3}$} & \multicolumn{1}{c|}{$s_i^{h_3}$} & \multicolumn{1}{c|}{$t_i^{h_3}$} \\
			\hline 
			\hline
			$h_3$-$h_3$-$h_3$ &
			\begin{tabular}[t]{@{}l@{}}
				$-\frac{3}{4}i \big[ 2c_\beta \mathcal{R}_{31} \left(2 \mathcal{R}_{31}^2\lambda_1 + 2 \mathcal{R}_{32}^2\lambda_{345} + \mathcal{R}_{33}^2\lambda_7 \right)v
				+ 2\mathcal{R}_{32} \left(2\mathcal{R}_{32}^2\lambda_2 + 2\mathcal{R}_{31}^2\lambda_{345} + \mathcal{R}_{33}^2\lambda_8\right) s_\beta v$ \\
				$+ \mathcal{R}_{33}\left(\mathcal{R}_{33}^2 \lambda_6 + 2\mathcal{R}_{31}^2 \lambda_7 + 2\mathcal{R}_{32}^2\lambda_8\right)v_s \big]$
			\end{tabular} & 1/2 & $i A_0(m_{h_3}^2)$ \\
			\hline
			$h_3$-$h_1$-$h_1$ & 
			\begin{tabular}[t]{@{}l@{}}
				$-\frac{1}{4}i \big[ 2 \left(2\mathcal{R}_{11} (2\mathcal{R}_{12}\mathcal{R}_{32}\lambda_{345} + \mathcal{R}_{13}\mathcal{R}_{33}\lambda_{7}) + \mathcal{R}_{31}(6\mathcal{R}_{11}^2\lambda_1 + 2\mathcal{R}_{12}^2\lambda_{345} + \mathcal{R}_{13}^2\lambda_7)\right)c_\beta v $ \\
				$+4\left( \mathcal{R}_{11}(2\mathcal{R}_{12}\mathcal{R}_{31} + \mathcal{R}_{11}\mathcal{R}_{32})\lambda_{345} + \mathcal{R}_{13}(\mathcal{R}_{12}\mathcal{R}_{33} + \frac{1}{2} \mathcal{R}_{13}\mathcal{R}_{32})\lambda_8 + 3\mathcal{R}_{12}^2\mathcal{R}_{32}\lambda_2\right)s_\beta v $ \\
				$+ \left(3\mathcal{R}_{13}^2\mathcal{R}_{33}\lambda_6 + 2\mathcal{R}_{11}(\mathcal{R}_{11}\mathcal{R}_{33} + 2\mathcal{R}_{13}\mathcal{R}_{31})\lambda_7 + 2\mathcal{R}_{12}(\mathcal{R}_{12}\mathcal{R}_{33} + 2\mathcal{R}_{13}\mathcal{R}_{32})\lambda_8\right)v_s \big]$ 
			\end{tabular} & 1/2 & $i A_0(m_{h_1}^2)$ \\    
			\hline
			$h_3$-$h_2$-$h_2$ & 
			\begin{tabular}[t]{@{}l@{}}
				$-\frac{1}{4}i \big[ 2 \left(6 \mathcal{R}_{21}^2\mathcal{R}_{31} \lambda_1 + 2\mathcal{R}_{22} (\mathcal{R}_{31} \mathcal{R}_{22} + 2\mathcal{R}_{21}\mathcal{R}_{32}) \lambda_{345} + \mathcal{R}_{23} ( \mathcal{R}_{31}\mathcal{R}_{23} + 2\mathcal{R}_{21}\mathcal{R}_{33}) \lambda_7 \right)c_\beta v $ \\
				$+4\left( \mathcal{R}_{21}(2\mathcal{R}_{22}\mathcal{R}_{31} + \mathcal{R}_{21}\mathcal{R}_{32})\lambda_{345} + \mathcal{R}_{23}(\mathcal{R}_{22}\mathcal{R}_{33} + \frac{1}{2} \mathcal{R}_{23}\mathcal{R}_{32})\lambda_8 + 3\mathcal{R}_{22}^2\mathcal{R}_{32}\lambda_2\right)s_\beta v $ \\
				$+ \left(3\mathcal{R}_{23}^2\mathcal{R}_{33}\lambda_6 + 2\mathcal{R}_{21}(\mathcal{R}_{21}\mathcal{R}_{33} + 2\mathcal{R}_{23}\mathcal{R}_{31})\lambda_7 + 2\mathcal{R}_{22}(\mathcal{R}_{22}\mathcal{R}_{33} + 2\mathcal{R}_{23}\mathcal{R}_{32})\lambda_8\right)v_s \big]$	  	
			\end{tabular} & 1/2 & $i A_0(m_{h_2}^2)$ \\
			\hline  
			$h_3$-$A$-$A$ & 
			\begin{tabular}[t]{@{}l@{}}
				$-\frac{1}{2}i \big[ 2c^3_\beta \mathcal{R}_{31}\left(\lambda_3+\lambda_4-\lambda_5\right)v + 2c_\beta \mathcal{R}_{31}\left(\lambda_1-2\lambda_5\right)s^2_\beta v 
				+ s^2_\beta \left(2\mathcal{R}_{32} (\lambda_3+\lambda_4-\lambda_5)s_\beta v + \mathcal{R}_{33}\lambda_7 v_s  \right) $ \\
				$+ c^2_\beta \left( 2\mathcal{R}_{32}(\lambda_2-2\lambda_5)s_\beta v + \mathcal{R}_{33} \lambda_8 v_s\right) \big]$           
			\end{tabular} & 1/2 & $i A_0(m_{A}^2)$ \\
			\hline              
			$h_3$-$G$-$G$ & 
			\begin{tabular}[t]{@{}l@{}}     
				$-\frac{1}{2}i \big[ 2c^3_\beta \mathcal{R}_{31} \lambda_1 v + 2c_\beta \mathcal{R}_{31} \lambda_{345} s^2_\beta v 
				+ s^2_\beta \left(2\mathcal{R}_{32} \lambda_2 s_\beta v + \mathcal{R}_{33}\lambda_8 v_s  \right)  
				+ c^2_\beta \left(2\mathcal{R}_{32} \lambda_{345} s_\beta v + \mathcal{R}_{33} \lambda_7 v_s\right) \big]$	   	                   
			\end{tabular} & 1/2 & $i A_0(\xi_Z m_Z^2)$ \\
			\hline  
			$h_3$-$H^+$-$H^-$ &
			\begin{tabular}[t]{@{}l@{}}
				$-\frac{1}{2}i \big[ 2c^3_\beta \mathcal{R}_{31} \lambda_3 v + 2c_\beta \mathcal{R}_{31} \left(\lambda_1-\lambda_4-\lambda_5\right) s^2_\beta v 
				+ s^2_\beta \left(2\mathcal{R}_{32} \lambda_3 s_\beta v + \mathcal{R}_{33}\lambda_7 v_s  \right) $ \\
				$+ c^2_\beta \left(2\mathcal{R}_{32} \left(\lambda_2-\lambda_4-\lambda_5\right) s_\beta v + \mathcal{R}_{33} \lambda_8 v_s\right) \big]$  	
			\end{tabular} & 1/2 & $2 \, \big[ i A_0(m_{H^\pm}^2)\big]$ \\
			\hline
			$h_3$-$G^+$-$G^-$ &
			\begin{tabular}[t]{@{}l@{}}
				$-\frac{1}{2}i \big[ 2c^3_\beta \mathcal{R}_{31} \lambda_1 v + 2c_\beta \mathcal{R}_{31} \lambda_{345} s^2_\beta v 
				+ s^2_\beta \left(2\mathcal{R}_{32} \lambda_2 s_\beta v + \mathcal{R}_{33}\lambda_8 v_s  \right) 
				+ c^2_\beta \left(2\mathcal{R}_{32} \lambda_{345} s_\beta v + \mathcal{R}_{33} \lambda_7 v_s\right) \big]$  	
			\end{tabular}  & 1/2 & $2 \, \big[i A_0(\xi_W m_W^2)\big]$ \\
			\hline
			$h_3$-$z$-$z$ & 
			\begin{tabular}[t]{@{}l@{}}
				$ ig m_W \left(c_\beta \mathcal{R}_{31} + s_\beta \mathcal{R}_{32} \right) / c^2_w $
			\end{tabular}  & 1/2 & $-i\big[(n-1) A_0(m_Z^2) + \xi_Z A_0(\xi_Z m_Z^2)\big]$ \\
			\hline
			$h_3$-$w^+$-$w^-$ &
			\begin{tabular}[t]{@{}l@{}}
				$ ig m_W \left(c_\beta \mathcal{R}_{31} + s_\beta \mathcal{R}_{32} \right)  $
			\end{tabular}  & 1/2 & $-2i \, \big[(n-1) A_0(m_W^2) + \xi_W A_0(\xi_W m_W^2)\big]$ \\
			\hline
			$h_3$-$u$-$\bar{u}$ & 
			\begin{tabular}[t]{@{}l@{}}
				$-\frac{i}{2} g \left(\mathcal{R}_{32}/ s_\beta \right) m_U / m_W   $  	
			\end{tabular}  & 1 & $-i \, m_{U} \, A_0(m_{U}^2) \,\text{Tr}(I_n)$  \\
			\hline
			$h_3$-$d$-$\bar{d}$ &
			\begin{tabular}[t]{@{}l@{}}
				$-\frac{i}{2} g \left(\mathcal{R}_{31}/ c_\beta \right) m_D / m_W   $
			\end{tabular}  & 1 & $-i \, m_{D} \, A_0(m_{D}^2) \,\text{Tr}(I_n)$   \\
			\hline
			$h_3$-$\eta_Z$-$\bar{\eta_Z}$ & 
			\begin{tabular}[t]{@{}l@{}}
				$ -\frac{i}{2} g m_W \left(c_\beta \mathcal{R}_{31} + s_\beta \mathcal{R}_{32} \right) \xi_Z  / c^2_w  $ 
			\end{tabular}  & 1 & $i \, A_0(\xi_Z m_Z^2)$   \\
			\hline
			$h_3$-$\eta_\pm$-$\bar{\eta_\pm}$ &
			\begin{tabular}[t]{@{}l@{}}
				$ -\frac{i}{2} g m_W \left(c_\beta \mathcal{R}_{31} + s_\beta \mathcal{R}_{32} \right) \xi_W   $          
			\end{tabular}  & 1 & $2i \, A_0(\xi_W m_W^2)$  \\
			\hline
		\end{tabular}
		\caption{The couplings $c_i^{h_3}$, symmetry factor $s_i^{h_3}$ and the propagator loops $t_i^{h_3}$ used for $T_{h_3}$ calculation.}
		\label{tab:tadpoleh3}
	\end{sidewaystable} 
	\clearpage
}

\clearpage
\bibliographystyle{apsrev4-2}
\bibliography{bibliography.bib}
\end{document}